\documentclass[10pt,onecolumn,floatfix,prb,superscriptaddress]{revtex4-2}
\usepackage{amsmath,amssymb,amsthm,mathrsfs,amsfonts,dsfont,amstext} 
\usepackage{soul} 
\usepackage{textcomp,pbox}
\usepackage[export]{adjustbox}
\usepackage{bm}
\usepackage{dcolumn,booktabs,url}
\usepackage[scaled]{helvet}
\usepackage{sansmath,gensymb}
\usepackage{tikz,graphicx,transparent,color}
\usepackage{multirow}
\usepackage[separate-uncertainty = true]{siunitx}
\usepackage{comment}
\usepackage{physics}
\usepackage{pdfpages}
\usepackage{array}
\usepackage{makecell}
\usepackage{dsfont}
\usepackage{tabularx}
\newcolumntype{L}{>{\raggedright\arraybackslash}X}
\newcolumntype{Y}{>{\centering\arraybackslash}X}

\makeatletter
\AtBeginDocument{\let\LS@rot\@undefined}
\makeatother

\newcolumntype{C}[1]{>{\centering\let\newline\\\arraybackslash\hspace{0pt}}m{#1}}

\usepackage{natbib} 

\usepackage[colorlinks=true]{hyperref}

\usepackage{mathtools} 

\hypersetup{
     colorlinks   = true,
     citecolor    = red,
     linkcolor    = blue,
     urlcolor     = red     
}

\graphicspath{{./}}

\newcommand{\SN}{Supplementary Note}

\newcommand{\ketUp}{{\ket{\Uparrow}}}
\newcommand{\ketDown}{{\ket{\Downarrow}}}

\newcommand{\ketupe}{{\ket{\uparrow\Downarrow\Uparrow}}}

\usepackage{lipsum}
\usepackage{graphicx}

\usepackage{import}
\usepackage{xifthen}
\usepackage{pdfpages}
\usepackage{transparent}
\usepackage{svg}
\usepackage{adjustbox}
\usepackage{calc}
\graphicspath{{./}}
\usepackage{physics}

\makeatletter
\def\maketitle{
\@author@finish
\title@column\titleblock@produce
\suppressfloats[t]}
\makeatother

\usepackage[resetlabels]{multibib}

\newcites{S}{References}

\begin{document}
\newcommand{\TitleName}{On-chip Spin-Photon Entanglement based on Photon-scattering of a Quantum Dot}
\title{\TitleName}

\newcommand{\AffCPH}{Center for Hybrid Quantum Networks (Hy-Q), The Niels Bohr Institute, University~of~Copenhagen,  DK-2100  Copenhagen~{\O}, Denmark}
\newcommand{\AffBasel}{Department of Physics, University of Basel, Klingelbergstra\ss e 82, CH-4056 Basel, Switzerland}
\newcommand{\AffBochum}{Lehrstuhl f\"ur Angewandte Fest\"orperphysik, Ruhr-Universit\"at Bochum, Universit\"atsstra\ss e 150, D-44801 Bochum, Germany}

\author{Ming Lai Chan}
\thanks{Email to: ming-lai.chan@nbi.ku.dk}
\affiliation{\AffCPH{}}

\author{Alexey Tiranov}
\thanks{Present address: Chimie ParisTech, Université PSL, CNRS, Institut de Recherche de Chimie Paris, 75005 Paris, France}
\affiliation{\AffCPH{}}

\author{Martin Hayhurst Appel}
\thanks{Present address: Cavendish Laboratory, University of Cambridge, JJ Thomson Avenue, Cambridge, CB3 0HE, UK}
\affiliation{\AffCPH{}}
\author{Ying Wang}
\affiliation{\AffCPH{}}
\author{Leonardo Midolo}
\affiliation{\AffCPH{}}
\author{Sven Scholz}
\affiliation{\AffBochum{}}
\author{Andreas D. Wieck}
\affiliation{\AffBochum{}}
\author{Arne Ludwig}
\affiliation{\AffBochum{}}
\author{Anders S{\o}ndberg S{\o}rensen}
\affiliation{\AffCPH{}}
\author{Peter Lodahl}
\affiliation{\AffCPH{}}

\date{\today}

\begin{abstract}
The realization of on-chip quantum interfaces between flying photons and solid-state spins is a key building block for quantum-information processors, enabling, e.g., distributed quantum computing, where remote quantum registers are interconnected by flying photons. Self-assembled quantum dots integrated in nanostructures are one of the most promising systems for such an endeavor thanks to their near-unity photon-emitter coupling and fast spontaneous emission rate. Here we demonstrate high-fidelity on-chip entanglement between an incoming photon and a stationary quantum-dot hole spin qubit. The entanglement is induced by sequential scattering of the time-bin encoded photon interleaved with active spin control within a microsecond, two-orders of magnitude faster than those achieved in other solid-state platforms. Conditioning on detection of a reflected photon renders the entanglement fidelity immune to spectral wandering of the emitter. These results represent a major step towards realizing a quantum node capable of interchanging information with flying photons and on-chip quantum logic, as required for quantum networks and quantum repeaters.
\end{abstract}

\maketitle 

In a future quantum network~\cite{Briegel2009}, remote quantum nodes could be connected by a large web of entangled photons. Traditionally these photonic states have been generated probabilistically by fusing smaller states, which typically requires an exponential overhead of ancillary photons~\cite{Brown2005}. The advent of a deterministic quantum interface between light and matter promises to radically change this notion \cite{Uppu2021}. For such systems, a flying photon is funneled into a nanophotonic structure and interacts efficiently with a quantum emitter that hosts a single spin~\cite{Lodahl2022}. Coherent manipulation of the spin state entangles it with the photon, forming the basis for deterministic quantum gates and, e.g., the generation of photonic cluster states for quantum computing \cite{Lindner2009,Pichler2017}.

So far, significant progress has been made towards this goal, particularly the realization of spin-photon entanglement~\cite{Togan2010,Gao2012,DeGreve2013,Schaibley2013,Schwartz2016,He2017,Nguyen2019,Appel2021b}, spin-spin entanglement~\cite{Delteil2016,Stockill2017,Nguyen2019,Daiss2021,Dordevic2021}, single-photon switching and swap gate~\cite{Sun2018,Javadi2018,Bechler2018}, and photon-photon entanglement~\cite{Hacker2016,Istrati2020,Thomas2022} using various quantum emitters. 
Among these platforms, quantum dots (QDs) integrated in nanophotonic structures offer near-unity coupling to light  ($\beta\geq98\%$)~\cite{Arcari2014} and a high photon-generation rate with near-unity purity and coherence~\cite{Tomm2021,Uppu2020}. Despite proven to be an excellent single-photon source, an interface capable of producing entanglement between a flying photon and the QD has yet to be demonstrated. This approach to entanglement generation is particularly attractive in the context of long-distance 
quantum communication, as the pulse profile of an incoming photon is preserved upon scattering~\cite{Duan2004}, which unlocks the prospects of interfacing distant emitters~\cite{Daiss2021}, realizing one-way quantum repeaters~\cite{Borregaard2020}, and performing quantum gate operations between two  photons~\cite{Hacker2016} or QDs~\cite{Mahmoodian2016}.

\begin{figure*}
	\includegraphics[width=0.9\linewidth]{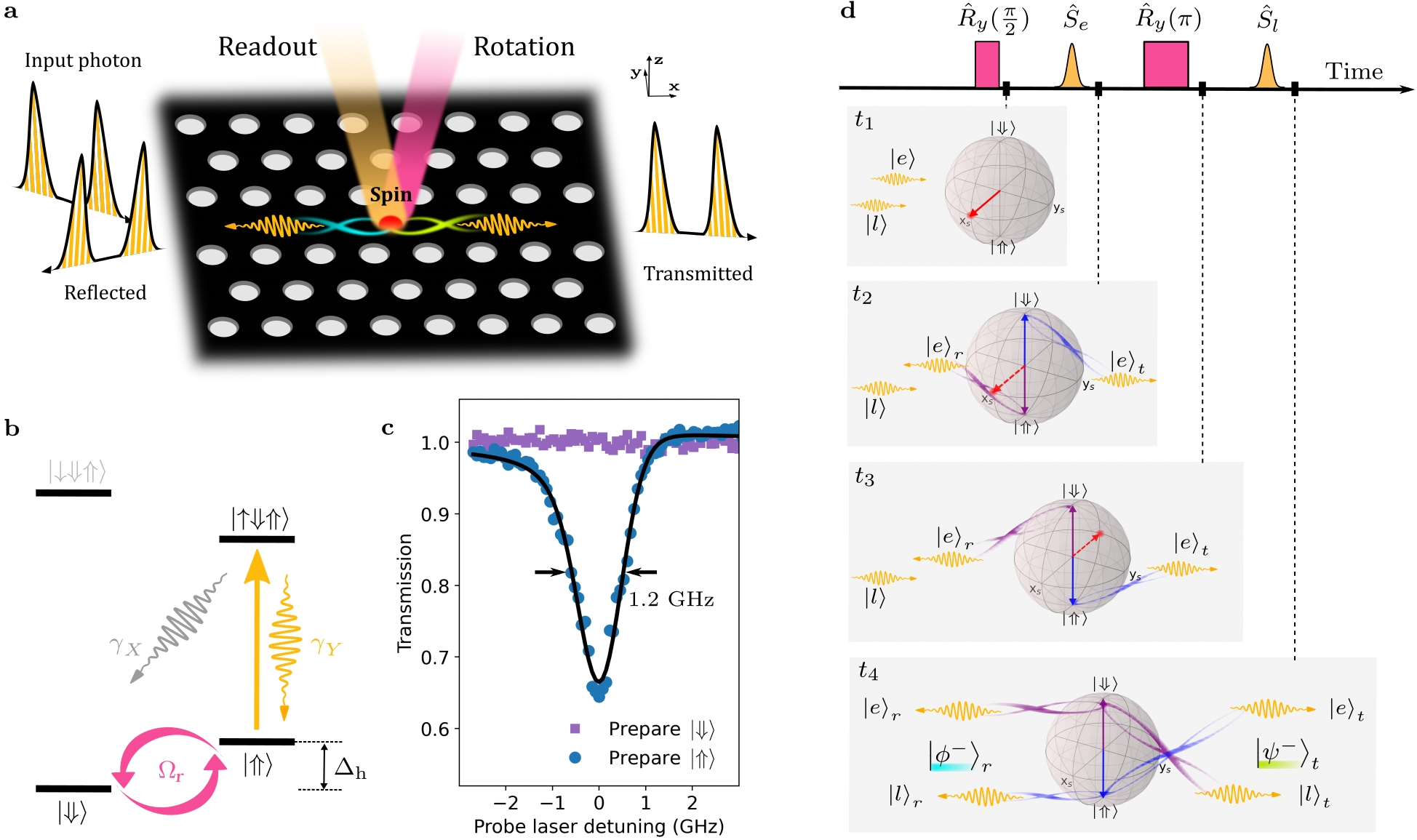}
	\caption{\textbf{Generation protocol of on-chip spin-photon entanglement.} \textbf{(a)} A coherently controlled spin in a QD (red) inside a photonic-crystal waveguide, where a Bell state (cyan lines) is generated upon conditional detection of a reflected photon. \textbf{(b)} QD level diagram. The excited state $\ketupe$ predominantly decays into $\ketUp$ with rate $\gamma_Y$ as $\gamma_Y\gg\gamma_X$. The wavelength of the main transition is 945~nm. Coherent control of the metastable hole spin ground states (magenta arrows, Rabi frequency $\Omega_r$) is realized via two-photon Raman processes by a detuned laser.
	\textbf{(c)} Single-photon transmission spectrum of the QD at $\textbf{B}_y=2$~T when preparing the spin state in either $\ketUp$ or $\ketDown$. \textbf{(d)} State evolution at different points in time during the protocol. At $t_1$, the QD spin (red) is prepared in a superposition state. At $t_2$, spin-dependent QD scattering occurs for the early time-bin $\ket{e}$. A $\pi$-rotation of the spin at $t_3$ is followed by scattering of the late time-bin $\ket{l}$ photon pulse at $t_4$. The two distinct Bell states $\ket{\phi^-}$ ($\ket{\psi^-}$) are generated conditioned on the detection of a reflected (transmitted) photon.}
	\label{fig:fig1}
\end{figure*}

\begin{figure*}
	\includegraphics[width=0.7\linewidth]{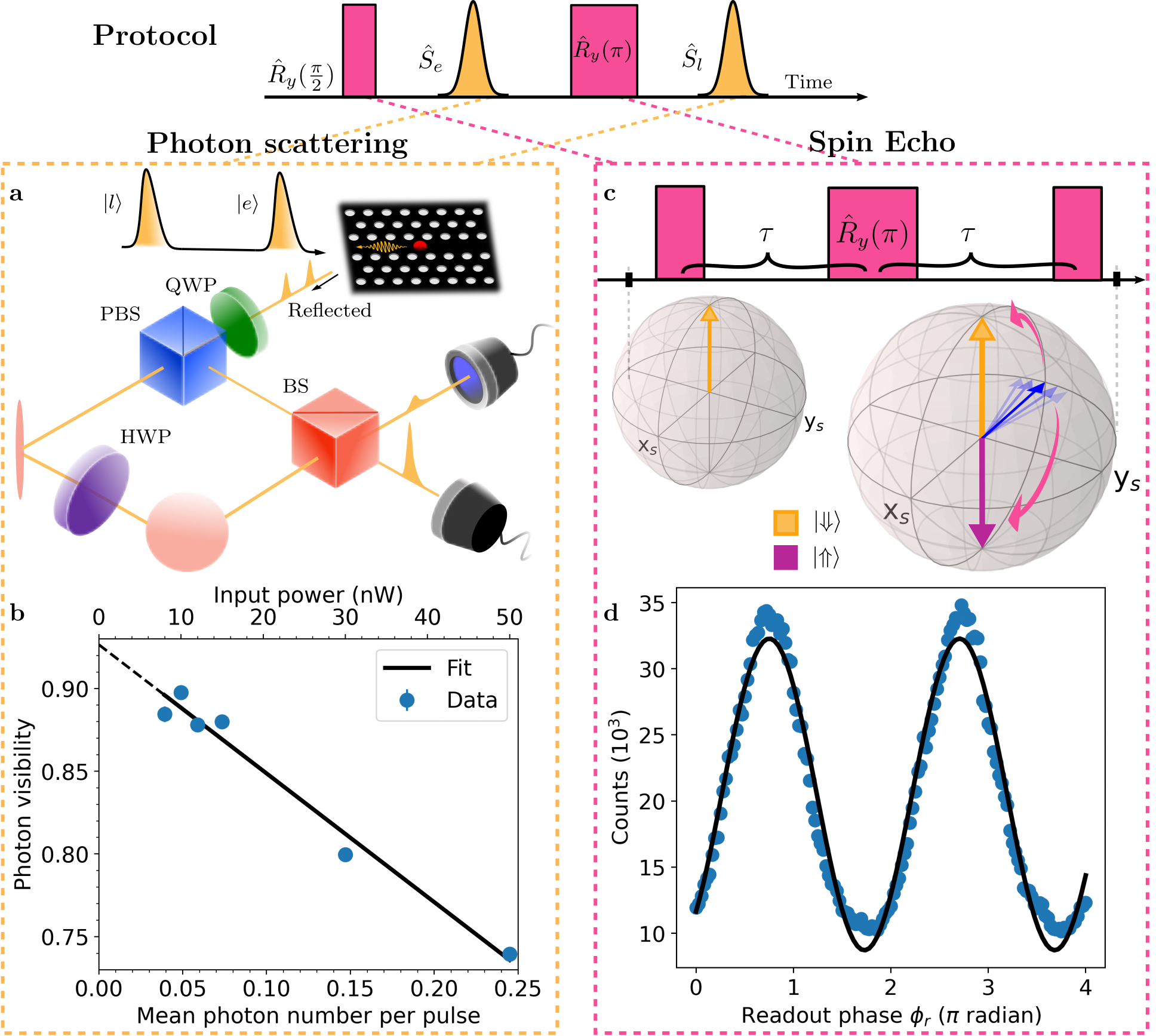} 
    
	\caption{\textbf{Coherent single-photon scattering and spin control.} \textbf{(a)} Setup for measuring photon visibility. The time-bin encoded qubit is reflected off the QD, which is initialized in the $\ket{\Uparrow}$ state. Upon entering the interferometer, the early time-bin is delayed which interferes with the late time-bin, constituting a $\ket{\pm X}_p=\ket{e}\pm\ket{l}$ basis measurement. PBS, polarizing beam-splitter; BS, 50:50 beam-splitter; QWP (HWP), quarter (half) wave-plate. 
	\textbf{(b)} Intensity visibility in photonic X-bases as a function of the mean photon number per pulse $\bar{n}$. The pure dephasing rate $\gamma_d$ is extracted from the $y$-intercept where $\bar{n}=0$.
	\textbf{(c)} Spin-echo sequence used to probe the spin coherence. The $\pi$-pulse is equally distant from the two $\pi/2$ pulses to eliminate inhomogeneous spin dephasing. The phase of the last $\pi/2$ pulse $\phi_r$ maps the equatorial state $\ketUp+e^{i\phi_r}\ketDown$ to the optically bright state $\ketUp$ ($\phi_r = \pi$) or dark state $\ketDown$ ($\phi_r = 0$).
	\textbf{(d)} Contrast between the spin $\ketDown$ and $\ketUp$ populations as a function of $\phi_r$. 
	}
	\label{fig:fig2}
\end{figure*}

\begin{figure*}
\includegraphics[width=0.8\linewidth]{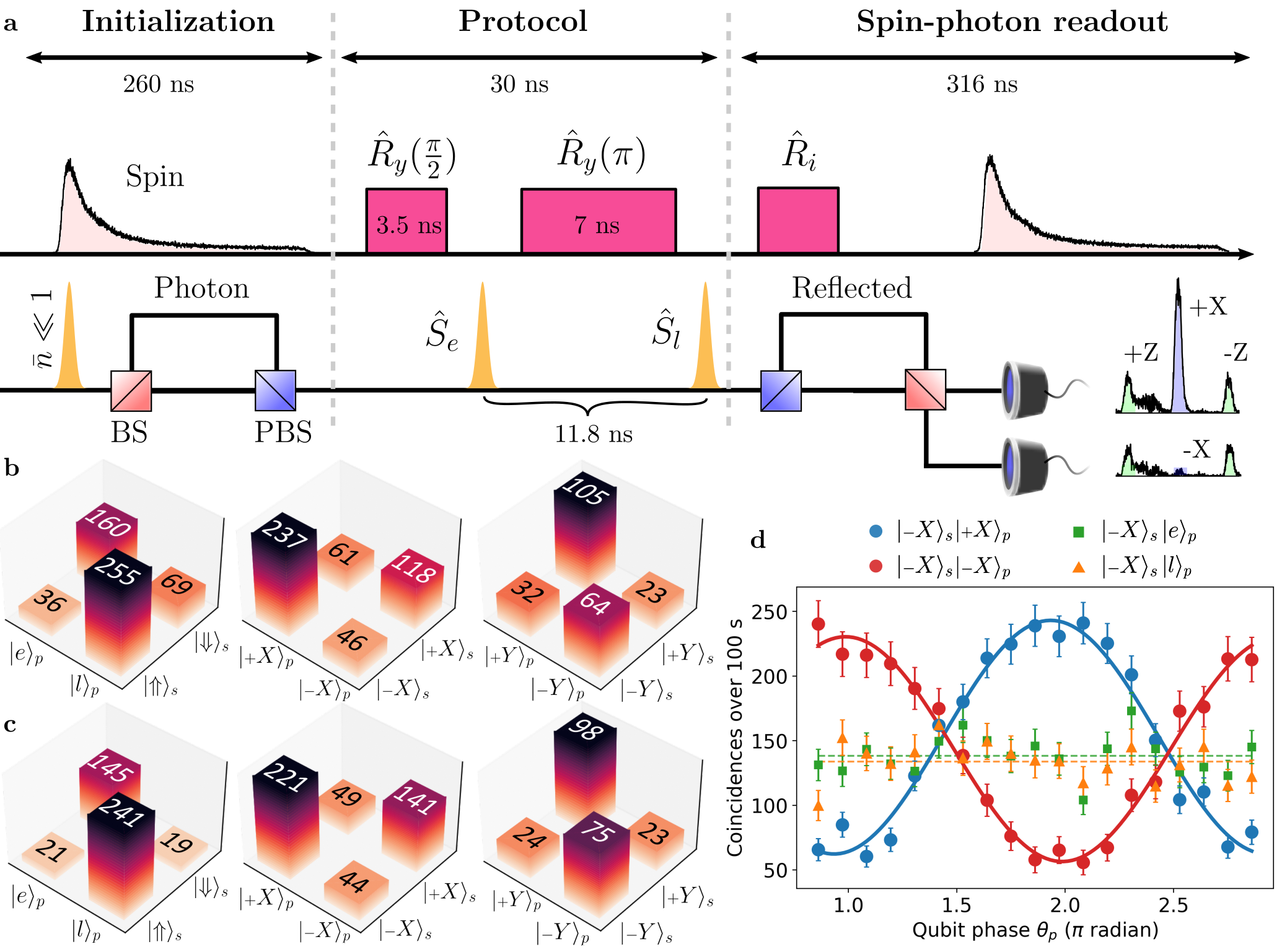}
	\caption{\textbf{Generation and verification of Bell states}. \textbf{(a)} Experimental sequence consisting of the preparation of spin and photonic qubits, the entanglement protocol and readout. The spin state is initialized and read out by optically driving the $\ketUp \leftrightarrow \ketupe$ transition (pale red), and $\hat{R}_i$ controls the spin projection basis. The photonic qubit is prepared and measured by the same interferometer in either the Z-basis (green) or equatorial basis (blue) time window.
	 \textbf{(b)} Raw two-photon coincidences measured during the photonic (p) readout window and spin (s) projections.
    \textbf{(c)} Measured two-photon coincidences corrected for laser background. \textbf{(d)} Visibility fringes of background-corrected two-photon coincidences as a function of the qubit phase $\theta_p$ when the spin state is projected on $\ket{-X}_s\propto\ket{\Uparrow}-\ket{\Downarrow}$. Circles (squares/triangles) correspond to projection on the photonic X-basis ($\pm$Z-basis). Solid curves are fits using $V_s\cos(\theta_p+\theta_{\text{offset}})$, and dashed lines are horizontal line fits.
	}
	\label{fig:fig3}
\end{figure*}

Here we step forward in this direction by demonstrating high-fidelity spin-photon entanglement between a guided photon and a QD spin embedded in a planar two-sided photonic-crystal waveguide. The entanglement is created in less than a microsecond, by sequential scattering of time-bin encoded photons using a QD heavy-hole spin. Conditioning on the detection of a reflected photon, renders the entanglement fidelity resistant to any residual spectral diffusion intrinsic to the emitter. The protocol can be extended to realize a fully deterministic entangling gate using single-sided waveguides~\cite{Borregaard2019}.
\newline\\
\noindent\textbf{Results}
\vspace{1mm}\newline
\textbf{Concept}. The protocol used to induce entanglement between a flying photon and the localized QD spin is outlined in Fig.~\ref{fig:fig1}. A single photon pulse is prepared in a superposition of an early $\ket{e}$ and a late $\ket{l}$ time-bin $\ket{\psi_p} = \alpha \ket{e} + \beta \ket{l}$ for $\alpha,\beta\in\mathbb{C}$ constituting a flying qubit. The photon is launched into a waveguide where the embedded QD spin is initialized in $\ket{\psi_s} = \ketDown$. The protocol proceeds by alternating between coherent spin rotations $\hat{R}_y$ and single-photon scattering $\hat{S}$, cf. Fig.~\ref{fig:fig1}d. A $\hat{R}_{y}(\pi/2)$ pulse prepares the spin in a superposition of the two spin ground states $\ketDown$ and $\ketUp$, while $\hat{R}_{y}(\pi)$ serves two purposes: (1) to invert the spin in-between the two scattering events to create entanglement, and; (2) to prolong the spin coherence time by acting as a spin-echo pulse between the two equally long time-bins~\cite{Tiurev2022}, when the spin is measured in the equatorial basis.  $\hat{S}$ corresponds to the photon being reflected (transmitted) when the QD state is in $\ketUp$ ($\ketDown$)~(Fig.~\ref{fig:fig1}c). When the flying photon is in an equatorial state, e.g., $\alpha=\beta=\frac{1}{\sqrt{2}}$, the ideal protocol results in the output state
\begin{align}
    \ket{\psi_{\text{out}}} &\propto [ \alpha \ket{e\Downarrow} - \beta \ket{l\Uparrow} ]_\textbf{r}
    +  [ \alpha \ket{e\Uparrow} - \beta \ket{l\Downarrow} ]_\textbf{t} \nonumber\\
    &\equiv \ket{\phi^-}_\textbf{r} + \ket{\psi^-}_\textbf{t},\label{eq:output}
\end{align}
which is a superposition of two spatially separated spin-photon Bell states. The subscript \textbf{r} (\textbf{t}) indicates that the photon was reflected (transmitted) by the QD. By post-selecting the detection on a photon being reflected (transmitted), the Bell state in the first (second) bracket is prepared. We find that the conditional Bell-state fidelity can approach unity for our system due to the spectral selectivity of the QD that predominantly reflects photons resonant with the transition $\ket{\Uparrow}\to\ket{\uparrow\Downarrow\Uparrow}$, despite residual QD spectral diffusion visible from the broadened transmission dip in Fig.~\ref{fig:fig1}c (\SN~4).
\newline\\
\noindent\textbf{Pre-calibration of the QD device}. We subdivide the entanglement protocol into two separate experiments~(Fig.~\ref{fig:fig2}). The first experiment probes the coherent nature of single-photon scattering (Fig.~\ref{fig:fig2}a), whereas the second experiment investigates the spin coherence with the built-in spin-echo sequence~(Fig.~\ref{fig:fig2}c). 
\newline\\
\textbf{Probing single-photon interference}. To demonstrate coherent scattering in the single-photon regime, we use a weak coherent state with a mean photon number per pulse $\bar{n} \ll1.$ We prepare a time-bin qubit using an asymmetric Mach-Zehnder interferometer where a photon is superposed between early and late temporal modes $\ket{e}$ and $\ket{l}$
, and scatter off the QD spin initialized in $\ket{\Uparrow}$. If the input photon of  spectral width $\sigma_o/2\pi$ is much narrower than the QD linewidth $\Gamma/2\pi$, the photon can be fully reflected due to destructive interference in transmission~\cite{Shen2005}. By interfering temporal modes of the reflected photon using the same interferometer and projecting on the X-basis $\ket{\pm X}_p=\ket{e}\pm\ket{l}$, the intensities $\text{I}_{\pm X}$ are measured, which are used to estimate the photon visibility $V_p\equiv \frac{\text{I}_{+X}-\text{I}_{-X}}{\text{I}_{+X}+\text{I}_{-X}}$. Due to the finite interferometric delay $\tau_{\text{delay}}=11.8$~ns, fluctuations occurring on longer timescales than  $\tau_{\text{delay}}$ (i.e., spectral diffusion of QDs~\cite{Kuhlmann2013}) are essentially filtered out, as they influence the reflected phase of both time-bins equally. The single-photon interference is thus subject to only fast dephasing processes. Notably, in the single-photon regime $(\bar{n}\approx0)$, we found $V_p=\frac{\Gamma}{\Gamma+2\gamma_d}$ for the total decay rate $\Gamma\approx2.48\text{ ns}^{-1}$~\cite{Appel2021b} and $\gamma_d$ is the pure dephasing rate (\SN~3). Here we measured a maximum value of $V_p=(89.7\pm0.4)\%$ which reduces linearly with $\bar{n}$~(Fig.~\ref{fig:fig2}b). Extrapolating a linear fit of $V_p$ to the $y$-intercept where $\bar{n}=0$ enables us to extract $\gamma_d\approx(0.099\pm0.004)\text{ ns}^{-1}$.
\newline\\
\textbf{Spin-Echo interferometry}. The second experiment benchmarks the coherence of the internal hole spin qubit. Specifically, we perform a spin-echo sequence~\cite{Hahn1950} consisting of two $\hat{R}_{y}(\pi/2)$ pulses separated by a $\hat{R}_{y}(\pi)$ pulse (Fig.~\ref{fig:fig2}c), which are implemented via the two-photon Raman scheme demonstrated in Ref.~\cite{Bodey2019} (see Methods). After the first $\hat{R}_{y}(\pi/2)$ pulse, due to fluctuating Overhauser nuclear fields~\cite{Urbaszek2013} the spin state begins to fan out over the Bloch sphere equator (denoted by blue arrows) decaying with a spin dephasing time $T^*_2=23.2$~ns~\cite{Appel2021b}. Applying a $\hat{R}_{y}(\pi)$ pulse after time $\tau$ inverts the direction of spin precession, thus refocusing the spin state at $t=2\tau$. The spin coherence is then probed by applying a second $\hat{R}_{y}(\pi/2)$ pulse, and scanning its phase $\phi_r$ followed by spin readout, which projects the resulting spin state onto either the optically bright or dark state. The resulting interferometric fringe is depicted in Fig.~\ref{fig:fig2}d with an extracted visibility of $V_s = (57.5\pm0.4)\%$ at $\tau=13$~ns, which is primarily limited by photo-induced incoherent spin processes~\cite{Appel2021b}. $V_s$ indicates how well the spin coherence is preserved, and benchmarks the quality of spin-photon correlations on the equatorial basis.
\newline\\
\textbf{Entanglement generation}. Having characterized the coherences of both qubits, we are in a position to run the entanglement protocol. The spin is first prepared in a superposition state $\ket{+X}_s \propto \ketUp + \ketDown$ by a $3.5$~ns $\hat{R}_{y}(\pi/2)$ pulse (Fig.~\ref{fig:fig3}a). The time-bin qubit is attenuated to $\Bar{n}\approx0.09$ before interacting with the QD (\SN~5). After sequential scattering of each time-bin, the reflected signal is collected and measured by the interferometer. Post-selecting the reflected photonic component carves out the output state [Eq.~(\ref{eq:output})] resulting in $\ket{\phi^-}_{\textbf{r}}$~\cite{Welte2017}. 

To determine the fidelity of the entangled state, we perform correlation measurements between the photonic modes and spin states. This involves projecting the entangled state on the $\hat{\sigma}_{i}^{(p)} \otimes \hat{\sigma}_{i}^{(s)} $ bases, where $i\in\{x,y,z\}$ denotes the Pauli operator, and the superscripts s (p) represent the spin and photonic qubits. The state of the reflected photon is detected in different time-bin windows after the interferometer, while the spin readout is performed by applying another rotation pulse $\hat{R}_i$ followed by optical driving of the main transition (See Methods; Fig.~\ref{fig:fig3}a). 

For each experimental setting, we condition on the detection of a reflected photon and the spin readout. The entanglement fidelity is measured using \cite{Appel2021b}
\begin{align}
    \mathcal{F}_{\text{Bell}} = \frac{\langle\hat{P}_z\rangle}{2} +\frac{\langle\hat{M}_y\rangle-\langle\hat{M}_x\rangle}{4}, 
\label{eq:Fe}
\end{align}
\noindent where $\langle\hat{M}_i\rangle=\langle\hat{\sigma}_{i}^{(p)} \otimes \hat{\sigma}_{i}^{(s)}\rangle$ is the normalized contrast, and $\langle\hat{P}_z\rangle\equiv(1+\langle\hat{M}_z\rangle)/2$. For measuring $\langle \hat{M}_{x/y} \rangle$, $\hat{R}_i=\hat{R}_{y/x}(\pi/2)$ is required for spin projection onto the equatorial state. Since the protocol now resembles a spin-echo sequence, the central $\hat{R}_{y}(\pi)$ pulse has an added benefit of spin-refocusing, whereas for Z-basis projections, spin echo is not necessary as $\langle\hat{P}_z\rangle$ is impervious to spin dephasing. As such, $|\langle \hat{M}_{x/y} \rangle|$ is dictated by the spin-echo visibility $V_s$, while $\langle\hat{P}_z\rangle$ largely reflects fidelity of the $\hat{R}_{y}(\pi)$ pulse $F_{\pi}$ (\SN~2). Figures~\ref{fig:fig3}b-d show the raw (background corrected) coincidence counts in various readout bases. We record $\langle\hat{P}_z\rangle = (90.7\pm2.2)\%$, $\langle\hat{M}_x\rangle=(-58.8\pm4.5)\%$ and $\langle\hat{M}_y\rangle=(57.3\pm6.6)\%$, where residual background counts from laser rotation pulses were subtracted (\SN~9). The recorded values of $|\langle \hat{M}_{x/y} \rangle|$ and $\langle \hat{P}_z \rangle$ are consistent with measured $V_s$ and $F_\pi$, respectively (see Methods). Using Eq.~(\ref{eq:Fe}), we obtain a corrected Bell-state fidelity of $\mathcal{F}_{\text{Bell}}= (74.3\pm2.3)\%$ (raw fidelity of $(66\pm2)\%$), which far exceeds the classical limit of $50\%$, clearly demonstrating the presence of entanglement in the generated quantum state.
\newline\\
\textbf{Theoretical modelling}. To unravel the physical mechanisms limiting the experimental fidelity, we derive the fidelity~(\SN~2) in the perturbative limit of small errors,
\begin{align}
    \mathcal{F}^{\text{theory}}_{\text{Bell}} \approx 1 - \frac{\gamma_d}{\Gamma} -\frac{\Gamma^2}{4\Delta_h^2}, 
\label{eq:F}
\end{align}
where $\Delta_h$ is the ground-state splitting. Eq.~(\ref{eq:F}) holds in the regime $\gamma_d \ll\Gamma \ll\Delta_h$ and assumes perfect manipulation of the hole spin state. In addition to being resilient to ground-state dephasing due to the built-in spin echo, the protocol is also impervious to errors arising from the spectral mismatch between the incoming optical pulse and the QD transition. This robustness is granted by the QD spectral reflectivity, which sifts out events where the photon interacts with the QD. The off-resonant frequency component of the incident pulse is transmitted without being detected thus having no impact on the entanglement fidelity.

Using Eq.~(\ref{eq:F}) with experimental parameters, the theoretical fidelity is estimated to be $\mathcal{F}^{\text{theory}}_{\text{Bell}} = 96.2\%$. Here the infidelity is attributed to decoherence from elastic phonon scattering~\cite{Tighineanu2018} $\gamma_d$ ($3.7\%$) and reflection from the off-resonant spin state $\Gamma/\Delta_h$ ($0.1\%$). The comparison to the experimental result indicates that several additional error mechanisms influence the experiment. The dominant cause is an incoherent photo-induced spin-flip error leading to non-ideal spin rotations, as is visible on the spin-echo data (Fig.~\ref{fig:fig2}d). These rotation errors together amount to a total infidelity of $15\%$ (\SN~2). Additional sources of error originate from driving-induced dephasing due to finite $\bar{n}$ ($7.2\%$) and imperfect spin readout ($2.7\%$), which are not intrinsic to the protocol. Taking these into account we estimate a theoretical lower fidelity bound $\mathcal{F}^{\text{theory}}_{\text{total}}\geq 73.0\%$, which agrees with the experimental value within the error margins.
Suppressing the photo-induced incoherent spin-flip processes is essential for improving the fidelity further. Encouragingly spin-rotation fidelities of $98.9 \%$ have been realized in the literature on electron spins~\cite{Bodey2019} and could be combined with nuclear-spin cooling methods to realize $T_2^*$ beyond 100~ns~\cite{Jackson2022}. With these improvements a near-unity entanglement fidelity is within reach.  
\newline
\\
\textbf{Discussions}
\vspace{1mm}\newline
The device performance is benchmarked by the entanglement fidelity, protocol speed and generation rate (\SN~7). In the present work, the demonstrated high entanglement fidelity ($74\%$) is competitive with previous solid-state implementations~\cite{Tchebotareva2019,Kim2013}, while the speed is improved. Indeed, the protocol operates on a sub-microsecond timescale ($0.6~\mu$s) which is at least 2 orders of magnitude faster than realized in SiV and atomic systems~\cite{Nguyen2019,Bhaskar2020,Kalb2015} as a consequence of the faster spin preparation time. The realized entanglement rate of $4.7$~Hz of the present device can be readily improved by increasing the collection efficiency and reflectivity of the device~(\SN~7).

Since the present implementation is conditioned on the detection of a reflected photon, the overall efficiency is bounded to at most 50\%. By adopting a single-sided waveguide or equivalently coupling the incident light to both reflection and transmission ports via a stabilized interferometer, a fully deterministic spin-photon quantum gate \cite{Borregaard2019} can eventually be realized. For a single-sided device all photons are reflected but with a spin-dependent phase. As such, no post-selection is required though the fidelity will be sensitive to spectral drifts of the QD transition to the second order. This however does not pose a fundamental limit to our platform as near-lifetime-limited QD transitions in photonic structures compatible with the proposed scheme were recently reported~\cite{Pedersen2020}.

We have demonstrated spin-photon entanglement by scattering an incoming photonic qubit off of a stationary QD spin. The system versatility is reflected by the fact that the same QD can also be operated as a source of multi-photon time-bin encoded entanglement generation~\cite{Appel2021b}. Such versatile spin-photon interfaces constitute building blocks of one-way quantum repeaters~\cite{Borregaard2020}. Furthermore, a range of new integrated quantum photonics devices and functionalities could potentially be realized, e.g.,  a deterministic Bell-state analyzer or a photonic quantum non-demolition detector \cite{Witthaut2012,Borregaard2019} that both rely on faithful coherent quantum state transfer from a flying photon to an emitter. The reflection-based scheme can be extended to realize non-local quantum entangling gates between distant quantum emitters~\cite{Daiss2021}. Finally, applying the above protocol interleaved with spin rotations in a single-sided device would realize entanglement between two subsequently incoming photons, i.e., a deterministic photon-photon quantum gate \cite{Hacker2016}, which is the most challenging quantum operation in photonics.
\newline
\\
\textbf{Methods}
\vspace{1mm}\newline
\textbf{Spin-photon interface}. To achieve the highly efficient light-matter interaction required by the protocol, we prepare a QD embedded in a suspended photonic-crystal waveguide (PCW) with two ports (\SN~1). The p-i-n heterostructure contains an intrinsic layer of self-assembled InAs QDs, enabling the electrical control of the QD charge state by applying a forward bias voltage. One experimental challenge is to simultaneously realize optical cycling transitions and spin control. This was recently achieved with a QD in a PCW under an in-plane magnetic field (Voigt geometry) by exploiting the inherent radiative asymmetry of the PCW \cite{Appel2021}. We employ a positively charged exciton giving access to a meta-stable hole spin ground state that was characterized in previous work \cite{Appel2021}. An in-plane external magnetic field ($\textbf{B}_y = 2$~T) Zeeman-splits the QD spin state into four energy levels, see Fig.~\ref{fig:fig1}b, where the linearly X- and Y-polarized dipoles form two $\Lambda$-systems. Thanks to the optical cyclicity of $C = \gamma_Y/\gamma_X \gg 1$ where the radiative decay rate $\gamma_Y$ ($\gamma_X$) is strongly enhanced (suppressed) by the PCW, an effective two-level system $\ket{\Uparrow}\leftrightarrow\ket{\uparrow\Downarrow\Uparrow}$ resembling a ``QD mirror" is realized. This leads to spin-dependent reflection of photons into the same frequency and polarization modes, granting the spectral selectivity necessary for the entanglement protocol. The relevant system rates and parameters are summarized in Table~1.
\vspace{1mm}\newline
\textbf{Experimental setup}. To perform high-fidelity entanglement experiments, the sample chip is cooled to 4.2 K inside a closed-cycle cryostat to suppress phonon scattering. A superconducting vector magnet provides a 2 T in-plane magnetic field enabling Raman transitions between two hole ground states. The sample is imaged with a 0.81 NA objective and brought to focus by translating 3 piezo positioners mounted beneath the sample. A DC voltage source provides a bias voltage at 1.148~V across the sample to populate QD charge states via tunnel coupling to a Fermi reservoir and control the charge environment. 

The experiment utilizes the same laser setup as in Ref.~\cite{Appel2021b} with a few notable differences: Two continuous wave (CW) lasers (linewidth $<10$~kHz) are used for the creation of the photonic qubit, resonant excitation of the QD and spin rotations. One of which is first directed to a double-pass acousto-optic modulator (AOM) setup followed by an electro-optical modulator
(EOM; iXBlue NIR-MX800-LN-20) to generate $2$~ns (FWHM) pulses for the photonic qubit. The non-diffracted light from the first AOM setup is then sent to a second AOM setup to create spin initialization and readout pulses ($200$~ns each) of the same laser frequency. The qubit laser pulses and QD emission are focused and collected at the same grating outcoupler using a cross-polarization scheme (\SN~1), while the readout laser is coupled directly on top of the QD (Fig.~\ref{fig:fig1}a). 

A photonic qubit encoded in time-bins is created by passing the 2~ns pulses through an asymmetric Mach-Zehner interferometer with a time delay of $\tau_{\text{delay}}=11.82$~ns. Here we chose the FWHM duration for the input pulse to be $2$ ns which exceeds the radiative lifetime of the optical transition $\Gamma^{-1} = 0.4$ ns for efficient single-photon scattering, but is narrow enough to be fitted within the $11.8$~ns time delay when combined with a $7$~ns $\pi$-rotation pulse and $1$~ns rise/fall time. The qubit phase $\theta_p$ can be scanned using a quarter-waveplate (QWP) and a linear polarizer. The reflected signal is then reinjected into the same interferometer and subsequently two narrowband ($3$~GHz) etalon filters to remove background from the rotation laser as well as QD phonon sidebands. The filtered signal then passes through a QWP and an EOM (not shown) which sets a 50/50 splitting ratio on the polarizing beam-splitter (see Fig.~\ref{fig:fig2}a). Since both the photonic qubit preparation and readout are performed via the same interferometer, the experiment becomes very robust against any mechanical or thermal drift allowing near-unity interferometric visibility on a week-long timescale~\cite{Appel2021b}.

Another CW laser is used for coherent spin control. It is sent through a third AOM setup and another EOM which is amplitude-modulated by a microwave source to create two sidebands with frequency difference matching the ground state splitting $\Delta_{h}/2\pi=7.3$ GHz, thus effectively driving the ground-state spin manifold. The sidebands are red-detuned from the cycling transition by $350$ GHz to avoid populating the excited states. The phase $\phi_r$ of the last microwave $\pi/2$-pulse is induced by a combination of a phase shifter and switches~\cite{Appel2021b} with a phase offset of $\sim0.3\pi$. The total pulse sequence duration is set to 606~ns. 
\newline\\
\textbf{Spin-photon state projections}. As shown in Fig.~\ref{fig:fig3}a, the detection of an early (late) photon traversing through the short (long) path of the interferometer constitutes the $\hat{\sigma}_{z}^{(p)}$-basis measurement (green). The spin readout in the $\hat{\sigma}_{z}^{(s)}$-basis is performed by applying another rotation pulse $\hat{R}_i=\hat{R}_y(0)$ ($\hat{R}_y(\pi)$) followed by optical driving of the main transition. Similarly, projection on the $\hat{\sigma}_{x}^{(p)} \otimes \hat{\sigma}_{x}^{(s)}$ ($\hat{\sigma}_{y}^{(p)} \otimes \hat{\sigma}_{y}^{(s)} $) bases is performed by detecting photons in the middle time window (blue) at $\theta_p\approx2\pi\equiv\theta_0$ ($\theta_p=\theta_0+\pi/2$) where the early and late time-bins between the short and long paths~\cite{Appel2021b} interfere, followed by $\hat{R}_i=\hat{R}_y(\pm\pi/2)$ ($\hat{R}_x(\pm\pi/2)$) before the spin readout. 
\newline
\\
\textbf{Data availability statement} The experimental data and analysis scripts of this study are available from the repository: \href{https://doi.org/10.17894/ucph.3b3bf28d-1f43-4b81-a03d-c9a9d3029b46}{https://doi.org/10.17894/ucph.3b3bf28d-1f43-4b81-a03d-c9a9d3029b46}.
\newline
\\
\textbf{Acknowledgements}
\vspace{1mm}\newline
The authors thank Yuxiang Zhang for fruitful discussions. We
gratefully acknowledge financial support from Danmarks
Grundforskningsfond (DNRF 139, Hy-Q Center for Hybrid Quantum Networks), Styrelsen for Forskning og Innovation (FI) (5072-00016B QUAN-TECH), the European Union’s Horizon 2020 research and innovation programme under grant agreement No. 820445 and project
name Quantum Internet Alliance, the European Union’s
Horizon 2020 Research and Innovation programme under
grant agreement No. 861097 (project name QUDOT-TECH). SS, ADW and AL gratefully acknowledge financial support from Deutsche Forschungsgemeinschaft
(DFG) (TRR 160 and LU2051/1-1) and the grants QR.X
16KISQ009 and DFH/UFA CDFA-05-06.
\newline
\\
\textbf{Competing interests} P.L. is founder of the company Sparrow Quantum that commercializes single-photon sources. The remaining authors declare no competing interests.
\newline
\\
\textbf{Author Contributions}
\vspace{1mm}\newline
M.L.C. and A.T. contributed equally to this work. M.L.C and A.T. performed the experiment, and theoretical analysis with help from A.S.S. Y.W. fabricated the device. A.L. and S.S. performed the heterostructure wafer growth with input from A.D.W. M.L.C and A.T. analyzed the data with help from M.H.A. M.L.C., A.T., and P.L wrote the manuscript with input from all authors. P.L., A.S.S., L.M., A.L., and A.D.W. supervised the project.
\bibliographystyle{naturemag} 
\let\oldaddcontentsline\addcontentsline
\renewcommand{\addcontentsline}[3]{}
\bibliography{reflist_clean}
\let\addcontentsline\oldaddcontentsline

\end{document}


\newcommand{\TitleName}{Supplementary Material for ``On-chip Spin-Photon Entanglement based on Photon-scattering of a Quantum Dot"}
\title{\TitleName}

\newcommand{\AffCPH}{Center for Hybrid Quantum Networks (Hy-Q), The Niels Bohr Institute, University~of~Copenhagen,  DK-2100  Copenhagen~{\O}, Denmark}
\newcommand{\AffBasel}{Department of Physics, University of Basel, Klingelbergstra\ss e 82, CH-4056 Basel, Switzerland}
\newcommand{\AffBochum}{Lehrstuhl f\"ur Angewandte Fest\"orperphysik, Ruhr-Universit\"at Bochum, Universit\"atsstra\ss e 150, D-44801 Bochum, Germany}

\author{Ming Lai Chan}
\thanks{Email to: ming-lai.chan@nbi.ku.dk; M.L.C. and A.T. contributed equally to this work.}
\affiliation{\AffCPH{}}

\author{Alexey Tiranov}
\thanks{Present address: Chimie ParisTech, Université PSL, CNRS, Institut de Recherche de Chimie Paris, 75005 Paris, France}
\affiliation{\AffCPH{}}

\author{Martin Hayhurst Appel}
\thanks{Present address: Cavendish Laboratory, University of Cambridge, JJ Thomson Avenue, Cambridge, CB3 0HE, UK}
\affiliation{\AffCPH{}}
\author{Ying Wang}
\affiliation{\AffCPH{}}
\author{Leonardo Midolo}
\affiliation{\AffCPH{}}
\author{Sven Scholz}
\affiliation{\AffBochum{}}
\author{Andreas D. Wieck}
\affiliation{\AffBochum{}}
\author{Arne Ludwig}
\affiliation{\AffBochum{}}
\author{Anders S{\o}ndberg S{\o}rensen}
\affiliation{\AffCPH{}}
\author{Peter Lodahl}
\affiliation{\AffCPH{}}

 
\maketitle
\onecolumngrid
\setcounter{equation}{0}
\setcounter{figure}{0}
\setcounter{table}{0}
\setcounter{page}{1}
\makeatletter

{
  \hypersetup{linkcolor=black}
  \tableofcontents
}
\setcounter{secnumdepth}{3} 
\renewcommand{\figurename}{\textbf{Supplementary Figure}}
\renewcommand{\tablename}{\textbf{Supplementary Table}}
\renewcommand{\refname}{Supplementary References}
\renewcommand{\thetable}{\arabic{table}}
\newpage
\section*{Supplementary Note 1: Cross-polarization Scheme on Waveguides}
\label{sec:setup}

\begin{figure}[h]
     \centering
     \includegraphics[scale=0.5]{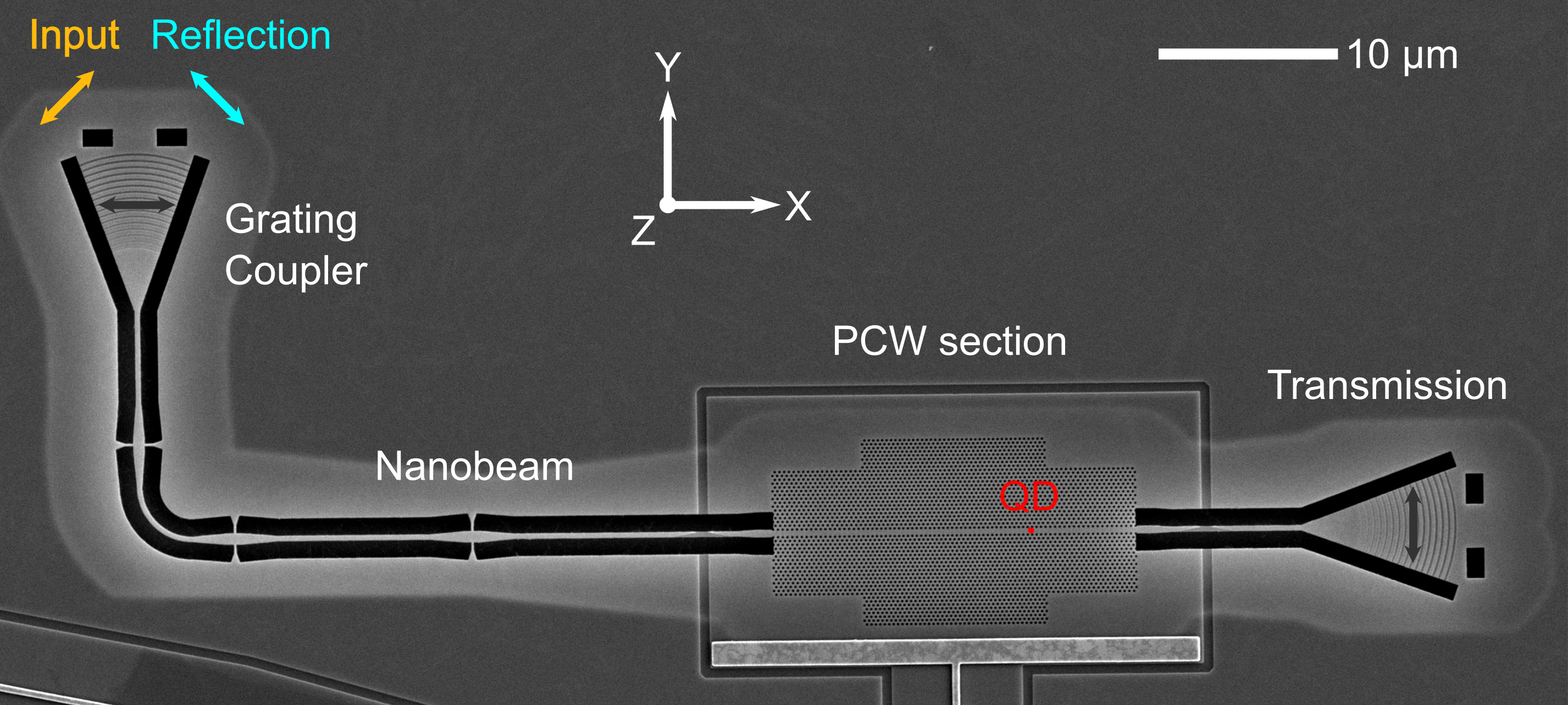}
     \caption{Scanning Electron Microscope (SEM) image of the waveguide and polarizations of the input and reflected light. Dark grey arrows denote the predefined polarizations of the grating couplers.}\label{fig:wg}
\end{figure}

The waveguide device used in the current experiment is terminated by two high-efficiency grating outcouplers~\cite{Zhou2018} representing reflection and transmission ports. For implementing the scattering experiment, both the incident laser and reflected signal are coupled to the same grating outcoupler (Supplementary Figure~\ref{fig:wg}). To distinguish the signal from the laser background, a cross-polarized scheme is used for excitation and collection. Before entering the waveguide, the polarization of the input light is carefully optimized with a set of waveplates such that it is orthogonal to the polarization of the collected light (set by another set of waveplates on the optical setup). The input is diagonally polarized (which can also be circularly polarized), thus only $50\%$ of the light is coupled to the grating coupler which has a predefined polarization along X~\cite{Zhou2018}. The X-polarized light is then converted into Y-polarization via the bend in the waveguide and subsequently interacts with the QD. Due to the non-chiral coupling of the waveguide, $50\%$ of the scattered signal thus returns to the same grating coupler, and further passes through the polarization control on the collection path resulting in another $50\%$ loss. Despite the loss in efficiency, the signal-to-noise ratio achieved in this setup can reach $\sim$100-300 depending on the mechanical stability of the optical setup. The transmission port constitutes a second collection path which can be used for both resonant transmission and entanglement experiments.
\newpage
\section*{Supplementary Note 2: Theory of Spin-photon Bell State Generation with Single-photon Scattering}
\label{sec:theory}
In this section, we discuss the state evolution of the spin-photon system upon applying the entanglement protocol and develop an analytical expression for the entanglement fidelity. Our strategy is similar to the approach taken in Ref.~\cite{Chan2022}, which is to first evaluate the fidelity to lowest order in perturbation theory for each of the considered errors. In the end the full fidelity is then found by multiplying the individual fidelities. 

We start with modeling a right-propagating time-bin photonic qubit $\alpha\ket{e}+\beta\ket{l}$ in a two-sided waveguide where $\alpha,\beta\in\mathbb{C}$, and the QD spin is initially in the ground state $\ket{\Downarrow}$. Following from Figure~1d of the main text, the entanglement protocol consists of (1) applying a $\hat{R}_y(\pi/2)$ spin rotation to prepare a superposition spin qubit, (2) scattering of the early photon $\hat{S_e}$, (3) a $\hat{R}_y(\pi)$-rotation, and finally (4) the scattering of the late photon $\hat{S_l}$. Each single-photon scattering process obeys the input-output relations~\cite{Witthaut2010}:
\begin{align}
    \underbrace{\ket{\omega \Uparrow}}_{\text{resonant}}\quad&\to\quad r_1 \ket{\omega \Uparrow}_r + t_1 \ket{\omega \Uparrow}_t + r_2 \ket{\omega_2 \Downarrow}_r + t_2 \ket{\omega_2 \Downarrow}_t; \nonumber\\
    \underbrace{\ket{\omega \Downarrow}}_{\text{off-resonant}}\quad&\to\quad \underbrace{\mathring{r}_1 \ket{\omega \Downarrow}_r}_{\substack{\text{reflection,} \\ \text{left-propagating}}} + \underbrace{\mathring{t}_1 \ket{\omega \Downarrow}_t}_{\substack{\text{transmission,} \\ \text{right-propagating}}} + \quad\underbrace{\mathring{r}_2 \ket{\omega-\Delta_h \Uparrow}_r}_{\substack{\text{reflection,} \\ \text{left, spin-flip}}}\quad + \quad\underbrace{\mathring{t}_2 \ket{\omega-\Delta_h \Uparrow}_t}_{\substack{\text{transmission,} \\ \text{right, spin-flip}}},\label{eq:sp}
\end{align}
where the photon in each time-bin is assumed to center around the resonant frequency of the dominant transition ($\ket{\Uparrow}\leftrightarrow\ket{\uparrow\Downarrow\Uparrow}$) with a Gaussian spectral profile, and $r_1$ ($t_1$) are the reflection (transmission) operators associated with the QD vertical transition $\ket{\Uparrow}\to\ketupe$ with a decay rate $\Gamma_1$ ($\gamma_Y$ in the main text). $r_2$ ($t_2$) corresponds to the diagonal transition $\ket{\Downarrow}\to\ketupe$ with decay rate $\Gamma_2$ ($\equiv\gamma_X$). $\omega_2=\omega+\Delta_h$ is the frequency of the Raman photon emitted from the diagonal transition where $\Delta_h$ is the ground-state splitting. The symbol $\circ$ denotes off-resonant scattering when the spin is in $\ket{\Downarrow}$. Below we use the superscript prime ($'$) to represent a scattered photon of frequency $\omega_2\neq\omega$. Using Eq.~(\ref{eq:sp}), the state evolution of the spin-photon system proceeds as
\begin{align}
    \bigg(\alpha\ket{e}+\beta\ket{l} \bigg)\otimes \ket{\Downarrow}&\xrightarrow{\hat{R}_y(\frac{\pi}{2})}\bigg(\alpha\ket{e}+\beta\ket{l} \bigg)\otimes \bigg(\ket{\Uparrow}+\ket{\Downarrow}\bigg)/\sqrt{2}\nonumber\\
    &\xrightarrow{\hat{S}_e} \bigg[ \alpha\bigg(r^e_1 \ket{e \Uparrow}_r + t^e_1 \ket{e \Uparrow}_t + r^e_2 \ket{e' \Downarrow}_r + t^e_2 \ket{e' \Downarrow}_t \bigg) \nonumber\\
    &\quad\quad+\alpha\bigg(\mathring{r}^e_1 \ket{e \Downarrow}_r + \mathring{t}^e_1 \ket{e \Downarrow}_t + \mathring{r}^e_2 \ket{e' \Uparrow}_r + \mathring{t}^e_2 \ket{e' \Uparrow}_t \bigg)+\beta\ket{l \Uparrow}+ \beta\ket{l \Downarrow} \bigg] /\sqrt{2}\nonumber\\
    &\xrightarrow{\hat{R}_y(\pi)} \bigg[ \alpha\bigg(-r^e_1 \ket{e \Downarrow}_r - t^e_1 \ket{e \Downarrow}_t + r^e_2 \ket{e' \Uparrow}_r + t^e_2 \ket{e' \Uparrow}_t \bigg) \nonumber\\
    &\quad\quad+\alpha\bigg(\mathring{r}^e_1 \ket{e \Uparrow}_r + \mathring{t}^e_1 \ket{e \Uparrow}_t - \mathring{r}^e_2 \ket{e' \Downarrow}_r - \mathring{t}^e_2 \ket{e' \Downarrow}_t \bigg)-\beta\ket{l \Downarrow}+ \beta\ket{l \Uparrow} \bigg] /\sqrt{2}\nonumber\\
    &\xrightarrow{\hat{S}_l} \bigg[ \alpha\bigg(-r^e_1 \ket{e \Downarrow}_r - t^e_1 \ket{e \Downarrow}_t + r^e_2 \ket{e' \Uparrow}_r + t^e_2 \ket{e' \Uparrow}_t \bigg) \nonumber\\
    &\quad\quad+\alpha\bigg(\mathring{r}^e_1 \ket{e \Uparrow}_r + \mathring{t}^e_1 \ket{e \Uparrow}_t - \mathring{r}^e_2 \ket{e' \Downarrow}_r - \mathring{t}^e_2 \ket{e' \Downarrow}_t \bigg)\nonumber\\
    &\quad\quad+\beta\bigg(r^l_1 \ket{l \Uparrow}_r + t^l_1 \ket{l \Uparrow}_t + r^l_2 \ket{l' \Downarrow}_r + t^l_2 \ket{l' \Downarrow}_t \bigg) \nonumber\\
    &\quad\quad-\beta\bigg(\mathring{r}^l_1 \ket{l \Downarrow}_r + \mathring{t}^l_1 \ket{l \Downarrow}_t + \mathring{r}^l_2 \ket{l' \Uparrow}_r + \mathring{t}^l_2 \ket{l' \Uparrow}_t \bigg)\bigg]/\sqrt{2}=\ket{\psi_{\text{out}}}.\label{eq:outstate}
\end{align}
In the ideal scenario where the early and late pulses are identical, monochromatic and resonant, and the QD optical cyclicity is infinite with no dephasing and loss, we have: (1) $r_1\to-1$ (resonant photons are coherently reflected with a $\pi$-phase shift), (2) $\mathring{t}_1\to1,\mathring{r}_1\to0$ (off-resonant photons are being transmitted instead of reflected), (3) $t_1\to0$ (complete destructive interference in the transmission); and (4) $r_2,\mathring{r}_2,t_2,\mathring{t}_2\to0$ (there are no Raman photons in the reflected and transmitted modes due to high cyclicity). As such, the ideal output state becomes $[ \alpha \ket{e\Downarrow} - \beta \ket{l\Uparrow} ]_\textbf{r}+  [ \alpha \ket{e\Uparrow} - \beta \ket{l\Downarrow} ]_\textbf{t}$. By preparing an equatorial photonic qubit $|\alpha|=|\beta|=\frac{1}{\sqrt{2}}$ and varying its phase $\theta_p$ where $\beta/\alpha= e^{i\theta_p}$, all 4 different Bell states can be generated upon conditioning on either the reflection or transmission of a scattered photon.
\begin{figure*}
	\includegraphics[width=0.4\linewidth]{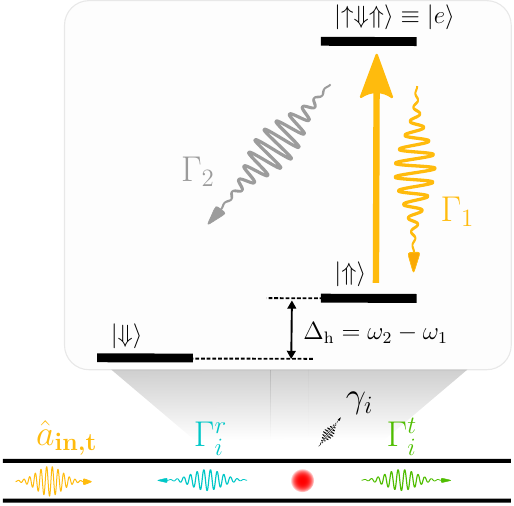}
	\caption{Level scheme for a QD embedded in a two-sided waveguide under the Voigt magnetic field. $\Gamma_1$ ($\Gamma_2$) is the radiative decay rate into the waveguide from the transition $\ket{e}\to\ket{\Uparrow}$ ($\ket{e}\to\ket{\Downarrow}$). $\Gamma_i=\Gamma^t_i+\Gamma^r_i$ for $i\in\{1,2\}$ includes both decay rates into the transmitted (`t") and reflected (`r") waveguide modes. $\gamma_i$ is the radiative rate into the lossy modes (Not to be confused with the radiative decay rates $\gamma_Y\equiv\Gamma_1$ and $\gamma_X\equiv\Gamma_2$ in the main text).}
	\label{fig:SM_QD}
\end{figure*}
\subsection{Scattering Coefficients for a \texorpdfstring{$\Lambda$}{Lg}-level Emitter in Two-sided Waveguides}
\label{subsec:sc3}
The scattering problem of a weak coherent state on the $\Lambda$-level emitter has been solved in Ref.~\cite{Das2018b} and its formalism can be easily extended to directly compute the scattering coefficients in Eq.~(\ref{eq:sp}). Specifically, the output field bosonic operator of the waveguide can be expressed in terms of the incident field and dynamical response of the emitter from the non-Hermitian Hamiltonian $\hat{\mathcal{H}}_{nh}$~\cite{Chan2022}. In a two-sided waveguide configuration, we label the field operator in the reflection port by the subscript ``r", and the transmitted port by ``t" (Supplementary Figure~\ref{fig:SM_QD}). Assuming that a right-propagating light field $\hat{a}^\dagger_{\text{in,t}}$ enters the waveguide, the output field operators on the transmitted (t) and reflected (r) ports are
\begin{align}
    &\ket{\omega \Uparrow}:
    \begin{dcases}
      \hat{a}_{\text{out,t}} & = \bigg[1-\frac{2\Gamma^t_1}{\Gamma+2i\delta_1}\hat{\sigma}_{11}-\frac{2\sqrt{\Gamma^t_1 \Gamma^t_2}}{\Gamma+2i\delta_1}\hat{\sigma}_{12} \bigg]\hat{a}_{\text{in,t}}\\
      \hat{a}_{\text{out,r}} & = \bigg[ -\frac{2\sqrt{\Gamma^t_1 \Gamma^r_1}}{\Gamma+2i\delta_1}\hat{\sigma}_{11}-\frac{2\sqrt{\Gamma^t_1 \Gamma^r_2}}{\Gamma+2i\delta_1}\hat{\sigma}_{12} \bigg]\hat{a}_{\text{in,t}}
    \end{dcases} \nonumber\\
    &\ket{\omega \Downarrow}:
    \begin{dcases}
      \hat{a}_{\text{out,t}} & = \bigg[1-\frac{2\Gamma^t_1}{\Gamma+2i(\delta_1+\Delta_h)}\hat{\sigma}_{11}-\frac{2\sqrt{\Gamma^t_1 \Gamma^t_2}}{\Gamma+2i(\delta_1+\Delta_h)}\hat{\sigma}_{12} \bigg]\hat{a}_{\text{in,t}}\\
      \hat{a}_{\text{out,r}} & = \bigg[ -\frac{2\sqrt{\Gamma^t_1 \Gamma^r_1}}{\Gamma+2i(\delta_1+\Delta_h)}\hat{\sigma}_{11}-\frac{2\sqrt{\Gamma^t_1 \Gamma^r_2}}{\Gamma+2i(\delta_1+\Delta_h)}\hat{\sigma}_{12} \bigg]\hat{a}_{\text{in,t}},\label{eq:intr}
    \end{dcases}
\end{align}
where $\delta_1=\omega_1 - \omega$ is the laser detuning from the transition $\ket{\Uparrow}\to\ket{e}$ for an emitter initialized in $\ket{\Uparrow}$. The total decay rate $\Gamma=\Gamma_1+\Gamma_2+\gamma_1+\gamma_2$ where $\Gamma_i$ ($\gamma_i$) is the radiative decay rate into (out of) the waveguide. $\Gamma_i=\Gamma^t_i+\Gamma^r_i$ includes both decay rates into the transmitted (`t") and reflected (`r") waveguide modes. $\Delta_h$ is the ground-state splitting. The output field operators have different detunings in their denominators because of different initial spin states of the QD: If the spin is initially $\ket{\Uparrow}$, the resonant frequency is $\omega_1$; If it is $\ket{\Downarrow}$ then the resonant frequency required to drive the diagonal spin transition is $\omega_2=\omega_1+\Delta_h$. $\hat{\sigma}_{ij}=\ket{j}\bra{i}$ is the atomic operator denoting a spin-flip in the atomic state when $i\neq j$. Note that when evaluating the probability of a spin-photon state, i.e., $\ket{e \Downarrow}_r$, the corresponding scattering coefficient $r^e_1(\omega)$ is first convoluted with a Gaussian lineshape $\Phi_1(\omega)$ and integrated with respect to $\omega$~\cite{Chan2022}. The individual resonant scattering coefficients in the frequency domain are
\begin{align}
    t_1(\omega) = 1-\frac{2\Gamma^t_1}{\Gamma+2i\delta_1},\quad \quad t_2(\omega) = -\frac{2\sqrt{\Gamma^t_1 \Gamma^t_2}}{\Gamma+2i\delta_1},\quad\quad r_1(\omega) = -\frac{2\sqrt{\Gamma^t_1 \Gamma^r_1}}{\Gamma+2i\delta_1},\quad\quad r_2(\omega) = -\frac{2\sqrt{\Gamma^t_1 \Gamma^r_2}}{\Gamma+2i\delta_1},\label{eq:scf}
\end{align}
where the off-resonant scattering coefficients are found similarly by replacing $\delta_1\to\delta_1+\Delta_h$.
\subsection{Projection Operators for Measuring Time-bin Encoded Photons}
At the end of the entanglement protocol, measurements to read out the state of the photonic qubit are performed by registering detector clicks in three different detection time windows. The detection of a time-bin photon is formulated by projection operators on different photonic readout bases:
\begin{align}
    \ket{e}\bra{e} = \int^\infty_{-\infty} \hat{a}^\dagger_e(t)\hat{a}_e(t) dt,\quad
    \ket{l}\bra{l} = \int^\infty_{-\infty} \hat{a}^\dagger_e(t+\tau)\hat{a}_e(t+\tau) dt,\quad
    \ket{e}\bra{l} = \int^\infty_{-\infty} \hat{a}^\dagger_e(t)\hat{a}_e(t+\tau) dt=(\ket{l}\bra{e})^\dagger,\label{eq:tb}
\end{align}
where the bosonic creation operator $\hat{a}^\dagger_e(t)$ represents the emission of a photon at time $t$ in the early time-bin, and $\tau$ is the interferometric delay. The projections $\ket{e}\bra{e}$ ($\ket{l}\bra{l}$) correspond to detecting photons in the side peak windows (green) (Figure~3a in the main text), whereas $\ket{e}\bra{l}$ refers to projection onto the middle detection window (blue central peak) where the early and late photons interfere. Since we only resolve the time-bin, the creation operator can be expressed in either the time or frequency domain. Using $a(t) = \frac{1}{\sqrt{2\pi}}\int^{\infty}_{-\infty}a(\omega)e^{i\omega t}d\omega$, one can show
\begin{align}
    \int^{\infty}_{-\infty}\hat{a}^\dagger_e(t) a_e(t) dt = \int^{\infty}_{-\infty}\hat{a}^\dagger_e(\omega)a_e(\omega)d\omega.\label{eq:fourier}
\end{align}
This implies we can adopt the same perturbation theory in the frequency domain to evaluate the fidelity as in Ref.~\cite{Chan2022}.

\subsection{Formula for the Operational Fidelity}
Now, with the time-bin projection operators defined, we can express the entanglement fidelity in terms of the above scattering coefficients. 
The measure of the quality of generated quantum states is conventionally given by the fidelity, which in our case evaluates the overlap between the output and ideal Bell states:
\begin{align}
    \mathcal{F}^{\text{theory}}_{\text{r}}&=\frac{\bra{\psi_{\text{ideal}}}\rho_{\text{out}}\ket{\psi_{\text{ideal}}}}{\Tr(\ket{\psi_{\text{out}}}\bra{\psi_{\text{out}}})}.\label{eq:ftheory}
\end{align}
Here the output reduced density matrix is given by $\rho_{\text{out}}=\Tr_{t,\omega} (\ket{\psi_{\text{out}}}\bra{\psi_{\text{out}}})$ which is a partial trace of the output density matrix $\ket{\psi_{\text{out}}}\bra{\psi_{\text{out}}}$ over the transmitted modes and frequency states $\omega\neq\omega_1$ not detected in the reflection. The total output density matrix $\rho_{\text{out}}$ is therefore obtained by effectively tracing out the unwanted modes. For simplicity we assume the use of perfect filters prior to detection which removes photons of frequencies other than $\omega_1$. The bandwidth of the etalon filters used in the experiment is $\sim3$ GHz with over $95\%$ transmission. This means the filter bandwidth is much narrower than the ground-state splitting $\Delta_h/2\pi=7.3$~GHz but wider than the QD homogeneous linewidth $\Gamma/2\pi=394$~MHz justifying the assumption. The fidelity in Eq.~(\ref{eq:ftheory}) is normalized by the success probability or efficiency $P_s=\Tr(\ket{\psi_{\text{out}}}\bra{\psi_{\text{out}}})\equiv\sum_i \bra{i}_r(\ket{\psi_{\text{out}}}\bra{\psi_{\text{out}}})\ket{i}_r$ since the protocol is conditioned on the detection of a photon in the reflection. In such case any event contributing to the loss of the scattered photon (e.g., finite cyclicity, nonzero coupling to leaky modes of the waveguide, and the transmission of a photon, which  is effectively treated as loss) does not reduce the fidelity. 

Using Eq.~(\ref{eq:outstate}), the normalized output reduced density matrix is found to be
\begin{align}
    \frac{\rho_{\text{out}}}{P_s} = \frac{1}{2P_s}\bigg(\abs{\alpha}^2 \abs{r^e_1}^2\ket{e \Downarrow}\bra{e \Downarrow} - \alpha\beta^* r^e_1 r^{l*}_1 \ket{e \Downarrow}\bra{l \Uparrow}- \alpha^* \beta r^{e*}_1 r^{l}_1 \ket{l \Uparrow}\bra{e \Downarrow} + \abs{\beta}^2 \abs{r^l_1}^2 \ket{l \Uparrow}\bra{l \Uparrow}+\abs{\alpha}^2 \abs{\mathring{r}^e_1}^2\ket{e \Uparrow}\bra{e \Uparrow}+...\bigg).\nonumber
\end{align}
For instance
we write out two of the matrix elements in $\rho_{\text{out}}$ using the results  from Sec.~\ref{subsec:sc3} and Eq.~(\ref{eq:fourier}):
\begin{align}
    \frac{1}{2}\abs{\alpha}^2\abs{r^e_1}^2 \ket{e \Downarrow} \bra{e \Downarrow} &= \frac{1}{2}\abs{\alpha}^2 \int^{\infty}_{-\infty}\int^{\infty}_{-\infty} d\omega d\omega' r_1(\omega) r^{*}_1(\omega') \Phi_1(\omega)\Phi_1(\omega') \hat{a}^\dagger_e (\omega) \ket{\emptyset \Downarrow}\bra{\emptyset \Downarrow} \hat{a}_e (\omega');\nonumber\\
    \frac{1}{2}\alpha\beta^* r^e_1 r^{l*}_1 \ket{e \Downarrow} \bra{l \Uparrow} &= \frac{1}{2}\alpha\beta^* \int^{\infty}_{-\infty}\int^{\infty}_{-\infty} d\omega d\omega' r_1(\omega) r^{*}_1 (\omega')\Phi_1(\omega)\Phi_1(\omega') \hat{a}^\dagger_e (\omega) \ket{\emptyset \Downarrow}\bra{\emptyset \Uparrow} \hat{a}_l (\omega'),\nonumber
\end{align}
where $\ket{\emptyset}$ is the vacuum state. For simplicity we now say that the early and late scattering events are identical for any given input frequency thus $r^e_1 = r^l_1 = r_1$. The reasoning behind this is further discussed in Sec.~\ref{subsec:perturbative}. Therefore, for an ideal Bell state in the reflected mode: $\ket{\psi_{\text{ideal}}}=\alpha\ket{e \Downarrow} - \beta\ket{l \Uparrow}$, the overlap of one of the density matrix elements with the ideal state becomes
\begin{align}
    &\bra{\psi_{\text{ideal}}} \bigg[ \frac{1}{2}\abs{\alpha}^2\abs{r^e_1}^2 \ket{e \Downarrow} \bra{e \Downarrow} \bigg] \ket{\psi_{\text{ideal}}} \nonumber\\
    &= \frac{\abs{\alpha}^4}{2}\int^{\infty}_{-\infty}\int^{\infty}_{-\infty}\int^{\infty}_{-\infty} d\omega d\omega' d\omega''\bra{\emptyset \Downarrow}\hat{a}_e(\omega'') \bigg[r_1 (\omega) r^{*}_1(\omega') \Phi_1(\omega) \Phi_1(\omega') \hat{a}^\dagger_e(\omega)\ket{\emptyset \Downarrow}\bra{\emptyset \Downarrow}\hat{a}_e(\omega')\bigg]\hat{a}^\dagger_e(\omega'') \ket{\emptyset \Downarrow}\nonumber\\
    &= \frac{\abs{\alpha}^4}{2}\int^{\infty}_{-\infty}\int^{\infty}_{-\infty}\int^{\infty}_{-\infty} d\omega d\omega' d\omega''r_1 (\omega) r^{*}_1(\omega') \Phi_1(\omega) \Phi_1(\omega') \delta(\omega-\omega'')\delta(\omega'-\omega'')\nonumber\\
    &= \frac{\abs{\alpha}^4}{2}\int^{\infty}_{-\infty} \abs{\Phi_1(\omega)}^2 \abs{r_1 (\omega)}^2 d\omega.\nonumber
\end{align}
Including all terms, the conditional  fidelity is found to be
\begin{align}
    \mathcal{F}^{\text{theory}}_{\text{r}}=\frac{1}{2P_s} \int^{\infty}_{-\infty} \abs{\Phi_1(\omega)}^2 \abs{r_1 (\omega)}^2 d\omega,\label{eq:of}
\end{align}
where the success probability $P_s$ is the trace of the output density matrix over the four basis states $\ket{i}=\{\ket{e \Uparrow}_r, \ket{e \Downarrow}_r, \ket{l \Uparrow}_r, \ket{l \Downarrow}_r \}$ in the Hilbert space of the spin-photon system. It is given by
\begin{align}
    P_s = \Tr(\ket{\psi_{\text{out}}}\bra{\psi_{\text{out}}})= \sum_i \bra{i}_r(\ket{\psi_{\text{out}}}\bra{\psi_{\text{out}}})\ket{i}_r= \frac{1}{2}\bigg[\int^{\infty}_{-\infty} \abs{\Phi_1(\omega)}^2\abs{r_1(\omega)}^2d\omega + \int^{\infty}_{-\infty} \abs{\Phi_1(\omega)}^2\abs{\mathring{r}_1(\omega)}^2d\omega\bigg].\label{eq:tr}
\end{align}
Combining Eqs.~(\ref{eq:of}) and (\ref{eq:tr}) results in the formula for the entanglement fidelity conditioned on reflected photons
\begin{align}
    \boxed{\mathcal{F}^{\text{theory}}_{\text{r}} = \frac{\int^{\infty}_{-\infty} \abs{\Phi_1(\omega)}^2\abs{r_1(\omega)}^2d\omega}{\int^{\infty}_{-\infty} \abs{\Phi_1(\omega)}^2\abs{r_1(\omega)}^2d\omega + \int^{\infty}_{-\infty} \abs{\Phi_1(\omega)}^2\abs{\mathring{r}_1(\omega)}^2d\omega}.}\label{eq:fc}
\end{align}

\subsection{Perturbative Form of the Entanglement Fidelity}
\label{subsec:perturbative}
The two integrals in Eq.~(\ref{eq:fc}) are the probabilities of scattering a photon of frequency $\omega_1$ from the spin state $\ket{\Uparrow}$ (resonant) and from $\ket{\Downarrow}$ (off-resonant) respectively. In particular, using Eq.~(\ref{eq:scf}) we find
\begin{align}
    \int^{\infty}_{-\infty}\abs{\Phi_1(\omega)}^2\abs{r_1(\omega)}^2 d\omega &=     \frac{1}{\sqrt{2\pi\sigma^2_o}}\int^{\infty}_{-\infty}e^{-\frac{(\omega-\omega_1)^2}{2\sigma^2_o}}\abs{-\frac{2\sqrt{\Gamma^t_1 \Gamma^r_1}}{\Gamma+2i\delta_1}}^2 d\omega \approx 1 - \frac{4\sigma^2_o}{\Gamma^2} - \frac{\Gamma^2-\Gamma^2_1}{\Gamma^2};\nonumber\\
    \int^{\infty}_{-\infty}\abs{\Phi_1(\omega)}^2\abs{\mathring{r}_1(\omega)}^2 d\omega &\approx \frac{\Gamma_1^2}{\Gamma^2+4\Delta^2_h},\label{eq:integral}
\end{align}
where we assume that  the scattered photon is equally coupled to the reflected and transmitted modes, i.e., $\Gamma^r_i=\Gamma^t_i = \Gamma_i/2$. $\sigma_o$ is the standard deviation of the spectral width of the incident Gaussian pulse. In evaluating Eq.~(\ref{eq:integral}) perturbatively we assume the frequency detuning $\delta_1$ to be small compared to the QD total decay rate $\Gamma$ and the ground-state splitting $\Delta_h$ for efficient light-matter interaction.

\subsubsection{Spectral mode mismatch}
 Conditioning on the detection of a reflected photon of frequency $\omega_1$ within the time-bin window, the entanglement fidelity becomes immune to the spectral error due to the nonzero bandwidth $\sigma_o$ of the incident pulse to lowest order in perturbation theory. Using Eqs.~(\ref{eq:fc}) and (\ref{eq:integral}), the resultant fidelity is
\begin{align}
    \mathcal{F}^{\text{theory}}_{\text{r}} \approx 1.
\end{align}
Simply stated, photons which are not resonant with the QD transition will be transmitted instead of reflected.
Since the protocol is conditioned on the reflection of either an early or a late photon, the transmission of the photon only reduces the success probability. The fidelity thus approaches unity as long as the dynamics of the early and late scattering events are identical. The same argument can be made for the broadening of the QD optical transition due to slow spectral wandering. Due to the ms-long spectral diffusion time~\cite{Kuhlmann2013}, the QD resonance drift is constant over its lifetime ($0.4~$ns) and also the interferometric delay ($11.8~$ns), the QD reflectivity is thus identical for both early and late scattering events without impacts on the entanglement fidelity. The spectral jittering on the QD resonance is modelled by taking $\delta_1=\omega_1-\omega\to\delta_1+\delta_e$ where $\delta_e$ follows a Gaussian spectral diffusion profile $N(0,\sigma_e)$~\cite{Chan2022}. 

If the protocol is post-selected on the presence of transmitted photon; however, the fidelity becomes susceptible to the spectral mismatch error. A similar analysis shows
\begin{align}
    \mathcal{F}^{\text{theory}}_{\text{t}} \approx 1-\frac{4\sigma^2_o}{\Gamma^2}-\frac{4\sigma^2_e}{\Gamma^2},
\end{align}
as the spectral infidelity arises from incomplete destructive interference between the incident field and the resonantly scattered photon ($t_1\neq0$). Any spectral effects reducing this interference would stain the quality of the entangled state. It is important to note that despite the QD spectral reflectivity, there is still a small probability of detecting undesired Raman photons of frequency $\omega_2 = \omega_1+\Delta_h$ in the reflection due to the finite optical cyclicity. These photons result from the imperfect QD two-level system and are filtered out.

\subsubsection{Finite cyclicity and coupling loss}
On the reflection port, photons could either originate from \textbf{(i)} resonant reflection on the spin-preserving transition (indicated by $r_1$), \textbf{(ii)} resonant Raman spin-flip process to $\ket{\Downarrow}$ ($r_2$), or \textbf{(iii)} off-resonant reflection from $\ket{\Downarrow}$ ($\mathring{r}_1$). A high cyclicity reduces the probability of resonant spin-flip process but strengthens off-resonant reflection. The undesired events (ii) and (iii) can be reduced by  having a larger ground-state splitting $\Delta_h\gg\Gamma$. Additionally, coupling to lossy modes of the waveguide implies that the reflected photons are lost without being detected; as a result these events do not affect the fidelity. Effectively we find
\begin{align}
    \mathcal{F}^{\text{theory}}_{\text{r}} &\approx 1-\frac{\Gamma^2}{4\Delta^2_h}\bigg(\frac{C}{C+1}\beta\bigg)^2.\label{eq:floss}
\end{align}
We observe that the fidelity is indeed robust to coupling loss and optical cyclicity, and is mainly reduced due to finite probability $\sim\Gamma^2/\Delta^2_h$ of detecting Rayleigh-scattered photons from $\ketDown$. Note that when deriving Eq.~(\ref{eq:floss}) we define the optical cyclicity $C\equiv\Gamma_1/\Gamma_2$~\cite{Appel2021}, the total decay rate $\Gamma=\Gamma_1+\Gamma_2+\gamma_1+\gamma_2$ where $\gamma_1$ ($\gamma_2$) is the radiative rate from the transition $\ket{e}\to\ket{\Uparrow}$ ($\ket{e}\to\ket{\Downarrow}$) which couples to lossy modes. The waveguide-coupling efficiency $\beta\equiv(\Gamma_1+\Gamma_2)/\Gamma$. From these conditions we obtain $\Gamma_1=\frac{C}{C+1}\beta\Gamma$ which is then substituted into Eq.~(\ref{eq:integral}).

\subsubsection{Phonon-induced pure dephasing}\label{subsubsec:pured}
The interaction of the QD with a phononic environment results in the broadening of the zero-phonon line and a broad phonon sideband~\cite{Besombes2001,Krummheuer2002,Muljarov2004,Tighineanu2018}. The latter can be filtered out while the former contributes to the reflection of incoherent photons which scramble the phase coherence of the spin-photon Bell state. The incoherent photons are only slightly broadened and thus cannot easily be removed by filters. 

We follow the approach in Ref.~\cite{Chan2022} and model this incoherent process as Markovian decoherence given by a dephasing rate $\gamma_d$ with the Lindblad operator $\sqrt{2\gamma_d}\hat{\sigma}_{ee}$ where $\ket{e}_s\equiv\ketupe$ is the atomic excited state. The dephasing leads to a quantum jump to the excited state (with a dephasing probability $P_{\gamma_d}$) followed by the decay to either of the two hole ground states with probabilities set by the transition rates $\Gamma_i/\Gamma$. The emitted photon into the waveguide is represented by a normalized photon density matrix $\rho^{\omega_i}_{\gamma_d}$. This is described by the density matrix
\begin{align}
    \rho' = \rho + P^{\omega_1}_{\gamma_d}\rho^{\omega_1}_{\gamma_d}\otimes\ket{\Uparrow}\bra{\Uparrow},\label{eq:1phonon}
\end{align}
where $\rho$ is the density matrix without a dephasing quantum jump. Initially there are also incoherent photons of frequency $\omega_2$ due to finite optical cyclicity but these are subsequently filtered out together with phonon sidebands. $\rho^{\omega_1}_{\gamma_d}\otimes\ket{\Uparrow}\bra{\Uparrow}$ is the photon density matrix resulting from the incoherent dephasing with a probability $P^{\omega_1}_{\gamma_d}$ given by
\begin{align}
    P^{\omega_1}_{\gamma_d} = \frac{\Gamma^r_1}{\Gamma} P_{\gamma_d}&= \frac{\Gamma^r_1}{\Gamma}\int^{\infty}_{-\infty} \frac{e^{-\frac{(\omega-\omega_1)^2}{2\sigma^2_o}}}{\sqrt{2\pi\sigma^2_o}}\abs{\frac{-2\sqrt{2\gamma_d\Gamma^r_1}}{\Gamma+2i(\omega-\omega_1)}}^2 d\omega=\frac{2\gamma_d}{\Gamma}\int^{\infty}_{-\infty} \abs{\Phi_1(\omega)}^2\abs{r_1(\omega)}^2d\omega.
    \label{eq:pured}
\end{align}
where $\Gamma^r_1=\Gamma_1/2$ is the decay rate in the reflected mode.

To evaluate the effect of pure dephasing in the entanglement protocol, it is instructive to consider the propagation of the error as there are two separate scattering events which will both lead to incoherent decay. Since Eq.~(\ref{eq:1phonon}) depends on whether there is a quantum jump to the excited state, we can assume that pure dephasing occurs primarily when the incident photon is resonant with the QD state since the excited state is unlikely to be populated via off-resonant scattering. As such, using Eq.~(\ref{eq:1phonon}) there are two additional incoherent density matrices in the normalized output reduced density matrix
\begin{align}
    \rho'_{\text{out}} = \frac{P_s \rho_{\text{out}} + \frac{1}{2}\abs{\alpha}^2 P^{\omega_1}_{\gamma_d}\rho^{\omega_1,e}_{\gamma_d}\otimes \hat{R}_y(\pi) \ket{\Uparrow}\bra{\Uparrow} \hat{R}^{\dagger}_y(\pi) + \frac{1}{2}\abs{\beta}^2 P^{\omega_1}_{\gamma_d}\rho^{\omega_1,l}_{\gamma_d}\otimes \ket{\Uparrow}\bra{\Uparrow} }{P_s + \Tr(\frac{1}{2}\abs{\alpha}^2 P^{\omega_1}_{\gamma_d}\rho^{\omega_1,e}_{\gamma_d}\otimes \hat{R}_y(\pi) \ket{\Uparrow}\bra{\Uparrow} \hat{R}^{\dagger}_y(\pi)) + \Tr(\frac{1}{2}\abs{\beta}^2 P^{\omega_1}_{\gamma_d}\rho^{\omega_1,l}_{\gamma_d}\otimes \ket{\Uparrow}\bra{\Uparrow})}.
\end{align}
Using Eq.~(\ref{eq:pured}) with $\abs{\alpha}=\abs{\beta}=1/\sqrt{2}$, the entanglement fidelity under pure dephasing is
\begin{align}
    \mathcal{F}^{\text{theory}}_{\text{r}} = \braket{\psi_{\text{ideal}}}{\rho'_{\text{out}}|\psi_{\text{ideal}}}&= \frac{\int^{\infty}_{-\infty} \abs{\Phi_1(\omega)}^2\abs{r_1(\omega)}^2d\omega+(\abs{\alpha}^4+\abs{\beta}^4)P^{\omega_1}_{\gamma_d}}{\int^{\infty}_{-\infty} \abs{\Phi_1(\omega)}^2\abs{r_1(\omega)}^2d\omega + \int^{\infty}_{-\infty} \abs{\Phi_1(\omega)}^2\abs{\mathring{r}_1(\omega)}^2d\omega+P^{\omega_1}_{\gamma_d}}\approx 1-\frac{\gamma_d}{\Gamma}.\label{eq:fidelity_pd}
\end{align}

\subsubsection{Spin dephasing}\label{subsubsec:spind}
In this section, we investigate how the decoherence of the spin states affects the entanglement fidelity. Specifically we consider the dephasing of the QD spin ground states, due to the presence of an external Overhauser field effectively formed by a neighboring nuclear ensemble~\cite{Urbaszek2013}. This effect causes a superposition spin qubit to precess on the equatorial plane at a random frequency $\delta_g$ slower than the QD decay rate, which can be modelled by applying a time-evolution operator $\hat{T}(\Delta t)=\exp(-i\delta_g \hat{S}_z \Delta t)$ on the superposition spin state, where $\hat{S}_z=\hat{\sigma}_z/2$~\cite{Chan2022}. During the entanglement sequence, a $\pi$-pulse is applied between two scattering events to ensure the precession of the spin is reversed and thus the spin is eventually refocused. In theory, the superposition qubit starts to precess at $t_0$ and the $\pi$-rotation pulse is applied at $t_\pi$. The spin is then refocused and read out at $t_r$ where $t_r-t_\pi=t_\pi-t_0=\Delta t$ must be satisfied for the perfect echo condition. For this model, the Overhauser field is assumed to be dominated by low-frequency nuclear noise $\omega\ll\frac{1}{2\Delta t}$ thus it is treated as quasi-static (``frozen"~\cite{Merkulov2002}) over the course of the experiment. The quantity of interest as a function of small drift $\delta_g$ can then be averaged with the Gaussian distribution $N(\delta_g,\sigma_{\text{OH}})$ where $\sigma_{\text{OH}}$ is the standard deviation in Overhauser field fluctuations.

To understand how spin echo benefits the entanglement protocol, we introduce the spin-echo operator $\hat{\mathcal{U}}_{\text{echo}} \equiv \hat{T}(t_r-t_\pi)\hat{R}_y(\pi)\hat{T}(t_\pi-t_0)$ which transforms the spin states into
\begin{align}
    \begin{cases}
      \hat{\mathcal{U}}_{\text{echo}}\ketUp  = -\exp(\frac{-i\delta_g (2t_\pi - t_r-t_0)}{2})\ketDown\equiv \lambda_{\Downarrow}\ketDown;\\
      \hat{\mathcal{U}}_{\text{echo}}\ketDown  = \exp(\frac{i\delta_g (2t_\pi - t_r-t_0)}{2})\ketUp\equiv\lambda_{\Uparrow}\ketUp.\label{eq:echo}
    \end{cases}
\end{align}
With Eq.~(\ref{eq:echo}), the normalized output state in Eq.~(\ref{eq:outstate}) becomes
\begin{align}
    \ket{\psi_{\text{out}}}=  -\alpha \lambda_{\Downarrow} r^e_1 \ket{e \Downarrow}_r +\alpha \lambda_{\Uparrow} \mathring{r}^e_1 \ket{e \Uparrow}_r +\beta \lambda_{\Uparrow} r^l_1 \ket{l \Uparrow}_r -\beta \lambda_{\Downarrow}\mathring{r}^l_1 \ket{l \Downarrow}_r +....\label{eq:sde}
\end{align}
Eq.~(\ref{eq:sde}) implies that the phase coherence between $\ket{e \Downarrow}_r$ and $\ket{l \Uparrow}_r$ depends on \textbf{(i)} the accumulated phase from spin precession, and \textbf{(ii)} the phase acquired from the early and late single-photon scattering events which is determined by the \textit{exact time} of scattering occurred within the optical pulse. Condition \textbf{(ii)} is made equal by interfering the time-bins with a matching time delay $\tau=11.8$~ns on the detection path. Since the time-bin qubit is created and measured using the same interferometer setup, by having an equal time delay $\tau_e=\tau_d=\tau$ for the excitation and detection paths, the interferometer temporally picks out events in which the exact time of scattering is in the same position of the pulse, i.e., $r^e_1(t')=r^l_1(t')$ for some time $t'\in \Phi_1(t)$ within the optical pulse. 

Now, to study how condition \textbf{(i)} affects the entanglement fidelity, we assume perfect single-photon scattering and consider only the output state conditioned on reflected photons, thus Eq.~(\ref{eq:sde}) is simplified as
\begin{align}
    \ket{\psi_{\text{out}}} = \alpha \lambda_{\Downarrow} \ket{e\Downarrow}_r-\beta \lambda_{\Uparrow} \ket{l\Uparrow}_r.\label{eq:zbasis}
\end{align}
For measuring the output state in the Z-basis, we compute the expectation value of the Z-basis projection operator $\hat{P}_z\equiv\ket{e\Downarrow}_r\bra{e\Downarrow}_r+\ket{l\Uparrow}_r\bra{l\Uparrow}_r$~\cite{Appel2021b,Guhne2007}
\begin{align}
    \langle\hat{P}_z\rangle \equiv \bra{\psi_{\text{out}}}\hat{P}_z\ket{\psi_{\text{out}}} = \abs{\alpha \lambda_{\Downarrow} }^2 + \abs{\beta \lambda_{\Uparrow}}^2 = 1,
\end{align}
which is insensitive to spin dephasing, regardless of whether the echo condition is fulfilled. However, when measuring in the X- (Y-basis), the corresponding expectation value of $\hat{M}_x$ ($\hat{M}_y$) averaged over $N(\delta_g,\sigma_{\text{OH}})$ is
\begin{align}
      \langle\hat{M}_x\rangle &=\int^{\infty}_{-\infty}\bra{\psi^x_{\text{out}}}\hat{M}_x\ket{\psi^x_{\text{out}}}N(\delta_g,\sigma_{\text{OH}})d\delta_g=-\int^{\infty}_{-\infty}\cos\frac{\delta_g\Delta\tau}{2}N(\delta_g,\sigma_{\text{OH}})d\delta_g=-e^{-(\Delta\tau/T^{*}_2)^2};\nonumber\\
      \langle\hat{M}_y\rangle &=\int^{\infty}_{-\infty}\bra{\psi^y_{\text{out}}}\hat{M}_y\ket{\psi^y_{\text{out}}}N(\delta_g,\sigma_{\text{OH}})d\delta_g=e^{-(\Delta\tau/T^{*}_2)^2},\label{eq:MxMy}
\end{align}
with the output state (Eq.~(\ref{eq:zbasis})) rewritten in the X- (Y-basis)
\begin{align}
      \ket{\psi^x_{\text{out}}} & =\frac{1}{2}\big[( \lambda_{\Downarrow}-\lambda_{\Uparrow})\ket{X^+_p X^+_s}-( \lambda_{\Uparrow}+\lambda_{\Downarrow})\ket{X^+_p X^-_s}+( \lambda_{\Uparrow}+\lambda_{\Downarrow})\ket{X^-_p X^+_s}+(\lambda_{\Uparrow}- \lambda_{\Downarrow})\ket{X^-_p X^-_s}\big];\nonumber\\
      \ket{\psi^y_{\text{out}}} & =\frac{1}{2}\big[-i( \lambda_{\Uparrow}+\lambda_{\Downarrow})\ket{Y^+_p Y^+_s}+i( \lambda_{\Uparrow}-\lambda_{\Downarrow})\ket{Y^+_p Y^-_s}+( \lambda_{\Downarrow}-\lambda_{\Uparrow})\ket{Y^-_p Y^+_s}+(\lambda_{\Downarrow}+ \lambda_{\Uparrow})\ket{Y^-_p Y^-_s}\big],\label{eq:outputxy}
\end{align}
and the respective projection operators
\begin{align}
    \hat{M}_x&=\ket{X^+_p X^+_s}\bra{X^+_p X^+_s}+\ket{X^-_p X^-_s}\bra{X^-_p X^-_s}-\ket{X^+_p X^-_s}\bra{X^+_p X^-_s}-\ket{X^-_p X^+_s}\bra{X^-_p X^+_s};\nonumber\\
    \hat{M}_y &=\ket{Y^+_p Y^+_s}\bra{Y^+_p Y^+_s}+\ket{Y^-_p Y^-_s}\bra{Y^-_p Y^-_s}-\ket{Y^+_p Y^-_s}\bra{Y^+_p Y^-_s}-\ket{Y^-_p Y^+_s}\bra{Y^-_p Y^+_s}.
\end{align}
Eq.~(\ref{eq:MxMy}) shows a Gaussian decay with spin dephasing time $T^*_2=\sqrt{2}/\sigma_{\text{OH}}$ when the echo condition  $\Delta\tau\equiv2t_\pi-t_r-t_0\neq0$ is not met. For that reason, to measure spin-photon correlations in the equatorial bases, a second $\hat{R}_y(\pi/2)$ pulse is applied at $t_r=2t_\pi-t_0$ to rotate the spin state to either of its two poles to prevent subsequent precession. From here, we note that for Z-basis fidelity measurements, the second $\hat{R}_y(\pi/2)$ pulse is not necessary as $\langle\hat{P}_z\rangle$ is tolerant to spin dephasing error. In such a case, the central $\pi$-rotation pulse does not play a refocusing role but is still required for inverting the spin between two scattering events. Additionally, due to having identical rotational pulse sequence, the spin-echo visibility $V_s$ measured in Figure~2d of the main text in principle establishes an upper bound for $|\langle\hat{M}_x\rangle|$ and $|\langle\hat{M}_y\rangle|$, while $\langle\hat{P}_z\rangle$ is primarily limited by fidelity of the $\hat{R}_y(\pi)$ pulse.

Note that Eq.~(\ref{eq:outputxy}) is derived from Eq.~(\ref{eq:zbasis}) taking $\beta=\alpha e^{i\theta_0}=\alpha$ for $\ket{\psi^x_{\text{out}}}$ ($\beta=\alpha e^{i(\theta_0+\pi/2)}=i\alpha$ for $\ket{\psi^y_{\text{out}}}$) (see Methods), where $\ket{e}\equiv\ket{X^+_p}+\ket{X^-_p}=\ket{Y^+_p}+\ket{Y^-_p}$ and $\ket{l}\equiv\ket{X^+_p}-\ket{X^-_p}=\ket{Y^+_p}-\ket{Y^-_p}$. The ideal case of Eq.~(\ref{eq:outputxy}) can be shown to be consistent with the experimental spin-photon correlations in Figures~3c-d (main text) taking $\Delta\tau=0$.

\subsubsection{Incoherent spin-flip error and finite \texorpdfstring{$T^*_2$}{}}
\label{subsec:spin-flip}
The next error concerns  spin decoherence induced by the red-detuned spin rotation laser and due to finite spin coherence time $T^*_2$. The former effect has been observed in Refs.~\cite{Bodey2019,Appel2021b} which results in power-dependent spin-flips, thereby destroying the coherence of the spin qubit during spin rotations. Despite its exact origin not being fully resolved, its effect on the spin coherence and the fidelity can be approximated by modelling the spin-flip error by a depolarizing channel $\mathcal{E}^s_{\text{depol}}$, with the probability of undergoing a random spin-flip $p$ dependent on the incoherent spin-flip rate $\kappa$ and the duration of the respective rotation pulse $T_r$. The action of the depolarizing channel on a density matrix $\rho$ is denoted by $\mathcal{E}_{\text{depol}}(\rho)=(1-p)\rho + p\mathcal{I}/2$, where $\mathcal{I}$ is the identity matrix. As an example, after applying a $\hat{R}_y(\pi/2)$ pulse on a spin state initialized in $\ket{\Downarrow}$, the spin density matrix transforms according to
\begin{align}
    \mathcal{E}^s_{\text{depol}}\bigg(\hat{R}^i_y(\pi/2) \rho_{\Downarrow} \hat{R}^{i\dagger}_y(\pi/2)\bigg)&=\mathcal{E}^s_{\text{depol}}\bigg(\mathcal{F}_{\frac{\pi}{2}}\hat{R}_y(\pi/2) \rho_{\Downarrow} \hat{R}^{\dagger}_y(\pi/2) + (1-\mathcal{F}_{\frac{\pi}{2}}) \rho_{-}\bigg)\nonumber\\
    &=(1-p_{\pi/2})\bigg(\mathcal{F}_{\frac{\pi}{2}}\hat{R}_y(\pi/2) \rho_{\Downarrow} \hat{R}^{\dagger}_y(\pi/2) + (1-\mathcal{F}_{\frac{\pi}{2}})\rho_{-} \bigg)+ \frac{p_{\pi/2}}{2}\mathcal{I}\nonumber\\
    &= 
    \left[\begin{array}{cc}
       \frac{1}{2} &  (1-p_{\pi/2})(\mathcal{F}_{\frac{\pi}{2}}-\frac{1}{2}) \\
        (1-p_{\pi/2})(\mathcal{F}_{\frac{\pi}{2}}-\frac{1}{2}) & \frac{1}{2}
    \end{array}\right]\equiv \mathbf{E}_{\pi/2},
\end{align}
where $\rho_{\Downarrow}$ is the initial spin density matrix and $\rho_{-} \equiv\ket{-}_s \bra{-}_s$. $\mathbf{E}_{\pi/2}$ is the output density matrix. In addition to the incoherent spin flip with a probability $p_{\pi/2}$ we here include known imperfections of the rotation pulse $\hat{R}^i_y(\pi/2)$, which has a fidelity of $\mathcal{F}_{\frac{\pi}{2}}$ to coherently rotate the spin to the superposition state $\ket{+}_s$ and a probability of $1-\mathcal{F}_{\frac{\pi}{2}}$ to project onto $\ket{-}_s$. The fidelity of \textit{coherent} $\pi/2$-spin rotation is determined by limitations of the two-photon Raman scheme, which is dominated by finite spin coherence time $T^*_2$ and the power-dependent rate $\gamma_r$ (Supplementary Note 6):
\begin{align}
    \mathcal{F}_{\frac{\pi}{2}} \approx (1-\gamma_r T_{\pi/2})\times \mathcal{F}_{\frac{\pi}{2}}(T^*_2),
\end{align}
where the $\pi/2$-rotation fidelity under the Overhauser field noise $\sigma_{\text{OH}}=\sqrt{2}/T^*_2$ is expressed by
\begin{align}
    \mathcal{F}_{\frac{\pi}{2}}(T^*_2) &\equiv \abs{\bra{\Downarrow}\hat{U}^\dagger_{rot,\text{ideal}}\hat{U}_{rot}\ket{\Downarrow}}^2 = \int^{\infty}_{-\infty} \frac{\Delta^2_{\text{OH}} + \Omega_r(\Omega_r+\sqrt{\Omega^2_r+\Delta^2_{\text{OH}}})}{2(\Omega^2_r+\Delta^2_{\text{OH}})}N(0,\sigma_{\text{OH}}) d\delta_g \approx 1 -\frac{2}{\pi^2}\bigg(\frac{T_{\pi/2}}{T_2^*}\bigg)^2,\label{eq:coherentF}
\end{align}
for a pulse duration of $T_{\pi/2}$. Eq.~(\ref{eq:coherentF}) is derived following the notations in Ref.~\cite{Appel2021thesis} where $\theta\equiv\frac{T_{\pi/2}}{2}\sqrt{\Omega^2_r+\Delta^2_{\text{OH}}}=\pi/4$, and $\Omega_r$ is the spin-rotation Rabi frequency with $\Omega_r T_{\pi/2}=\pi/2$. $\mathcal{F}_{\pi}$ can also be derived similarly. The probability of introducing a depolarizing error $p_{\pi/2}$ during a $\hat{R}_y(\pi/2)$ rotation is estimated by integrating the exponential distribution over the pulse duration for a given incoherent spin-flip rate $\kappa$:
\begin{align}
    p_{\pi/2}=\int^{T_{\pi/2}}_0 \kappa e^{-\kappa t}dt=1-e^{-\kappa T_{\pi/2}}.\label{eq:pih}
\end{align}
The exponential distribution describes the probability of a random spin-flip occurring in a certain time period, where the spin-flip event is assumed not to  depend on how much time has passed in the protocol (i.e. it is memory-less). Similarly, for a $\hat{R}_y(\pi)$ pulse applied on an arbitrary spin state $\rho_s$,
\begin{align}
    \mathcal{E}^s_{\text{depol}}\bigg(\hat{R}^i_y(\pi) \rho_s \hat{R}^{i\dagger}_y(\pi)\bigg)&=(1-p_\pi)\bigg(\mathcal{F}_\pi\hat{R}_y(\pi) \rho_s \hat{R}^{\dagger}_y(\pi) + (1-\mathcal{F}_\pi)\rho_s \bigg)+ \frac{p_\pi}{2}\mathcal{I}\nonumber\\
    &= 
    \left[\begin{array}{cc}
       (1-p_\pi)[\mathcal{F}_\pi \rho_4 + (1-\mathcal{F}_\pi)\rho_1] +\frac{p_\pi}{2}  &  (1-p_\pi)[-\mathcal{F}_\pi \rho_3 + (1-\mathcal{F}_\pi)\rho_2] \\
        (1-p_\pi)[-\mathcal{F}_\pi \rho_2 + (1-\mathcal{F}_\pi)\rho_3] & (1-p_\pi)[\mathcal{F}_\pi \rho_1 + (1-\mathcal{F}_\pi)\rho_4] + \frac{p_\pi}{2}
    \end{array}\right]\equiv \mathbf{E}_{\pi}\label{eq:depolpi},
\end{align}
where the initial spin density matrix is
\begin{align}
    \rho_s\equiv\left[\begin{array}{cc}
       \rho_{1}  &  \rho_{2} \\
        \rho_{3} & \rho_{4}
    \end{array}\right],
\end{align}
and $p_\pi$ is the probability of introducing the depolarizing error during a $\hat{R}_y(\pi)$ rotation found similarly as in Eq.~(\ref{eq:pih}). As a quick sanity check, using Eq.~(\ref{eq:depolpi}) and $\rho_1=\rho_2=\rho_3=0,\text{ }\rho_4=1$, the total $\pi$-rotation pulse fidelity which includes the contribution from both coherent and incoherent spin-flip processes can be estimated to be
\begin{align}
    \mathcal{F}_{\pi,\text{total}} &= (1-p_\pi)\mathcal{F}_\pi + \frac{p_\pi}{2}= (1-p_\pi)\times(1-\gamma_r T_\pi)\times\mathcal{F}_\pi(T^*_2) + \frac{p_\pi}{2}\approx 1-\frac{1}{2}(\kappa+2\gamma_r) T_{\pi} -\frac{2}{\pi^2}\bigg(\frac{T_{\pi}}{T^*_2}\bigg)^2.
\end{align}
Using experimental values for the incoherent spin-flip rate $\kappa=0.0098\text{ ns}^{-1}$ and $\gamma_r=0.0081\text{ ns}^{-1}$ extracted in Supplementary Note 6 with spin dephasing time $T^*_2=23.2\text{ ns}$~\cite{Appel2021b}, we estimate $\mathcal{F}_{\pi,\text{total}}\approx 89.8\%$ for $T_{\pi}=7$~ns, which indeed agrees with the measured value of $F_\pi=(88.1\pm3.8)\%$ (Supplementary Note 6).
\newpage
Now we consider the evolution of the spin-photon system during the entanglement protocol. The protocol begins by preparing a time-bin photonic qubit $\rho_p$ and a spin state in $\rho_s$:
\begin{align}
    \rho_p \otimes \rho_s
    =& \left[\begin{array}{cc}
       \abs{\alpha}^2  &  \alpha^*\beta \\
        \alpha\beta^* & \abs{\beta}^2
    \end{array}\right] \otimes \left[\begin{array}{cc}
       0  &  0 \\
        0 & 1
    \end{array}\right]\nonumber\\
    \xrightarrow{\hat{R}_y(\pi/2)}&\quad (\mathcal{I}\otimes \mathcal{E}^s_{\text{depol}}) ( \rho_p \otimes \rho_s)
    = \left[
        \begin{array}{@{}cc@{}}
          \abs{\alpha}^2\mathbf{E}_{\pi/2}
          & \alpha^*\beta\mathbf{E}_{\pi/2} \\
          \alpha\beta^*\mathbf{E}_{\pi/2} &
          \abs{\beta}^2\mathbf{E}_{\pi/2}
        \end{array}
        \right]
    \equiv\left[
        \begin{array}{@{}cc@{}}
          \abs{\alpha}^2\left[\begin{matrix}
          E^1_{\pi/2} & E^2_{\pi/2} \\
          E^3_{\pi/2} & E^4_{\pi/2}
          \end{matrix}\right]
          & \alpha^*\beta\left[\begin{matrix}
          E^1_{\pi/2} & E^2_{\pi/2} \\
          E^3_{\pi/2} & E^4_{\pi/2}
          \end{matrix}\right] \\
          \alpha\beta^*\left[\begin{matrix}
          E^1_{\pi/2} & E^2_{\pi/2} \\
          E^3_{\pi/2} & E^4_{\pi/2}
          \end{matrix}\right] &
          \abs{\beta}^2\left[\begin{matrix}
          E^1_{\pi/2} & E^2_{\pi/2} \\
          E^3_{\pi/2} & E^4_{\pi/2}
          \end{matrix}\right]
        \end{array}
        \right]\nonumber\\
    \xrightarrow{\hat{S}_e}&\quad
        \left[\begin{array}{@{}cc@{}}
          \abs{\alpha}^2\left[\begin{matrix}
          \abs{r^e_1}^2 E^1_{\pi/2} & r^{e*}_1 \mathring{r}^{e}_1 E^2_{\pi/2} \\
          r^e_1 \mathring{r}^{e*}_1 E^3_{\pi/2} & \abs{\mathring{r}^{e}_1}^2 E^4_{\pi/2}
          \end{matrix}\right]
          & \alpha^*\beta\left[\begin{matrix}
          r^{e*}_1 E^1_{\pi/2} & r^{e*}_1 E^2_{\pi/2} \\
          \mathring{r}^{e*}_1 E^3_{\pi/2} & \mathring{r}^{e*}_1 E^4_{\pi/2}
          \end{matrix}\right] \\
          \alpha\beta^*\left[\begin{matrix}
          r^e_1 E^1_{\pi/2} & \mathring{r}^e_1 E^2_{\pi/2} \\
          r^e_1 E^3_{\pi/2} &\mathring{r}^e_1 E^4_{\pi/2}
          \end{matrix}\right] &
          \abs{\beta}^2\left[\begin{matrix}
          E^1_{\pi/2} & E^2_{\pi/2} \\
          E^3_{\pi/2} & E^4_{\pi/2}
          \end{matrix}\right]
        \end{array}
        \right]
    \equiv\left[\begin{array}{@{}cc@{}}
          \abs{\alpha}^2\left[\begin{matrix}
          \rho^1_1 & \rho^2_1 \\
          \rho^3_1 & \rho^4_1
          \end{matrix}\right]
          & \alpha^*\beta\left[\begin{matrix}
          \rho^1_2 & \rho^2_2 \\
          \rho^3_2 & \rho^4_2
          \end{matrix}\right] \\
          \alpha\beta^*\left[\begin{matrix}
          \rho^1_3 & \rho^2_3 \\
          \rho^3_3 & \rho^4_3
          \end{matrix}\right] &
          \abs{\beta}^2\left[\begin{matrix}
          \rho^1_4 & \rho^2_4 \\
          \rho^3_4 & \rho^4_4
          \end{matrix}\right]
        \end{array}
        \right]
        \nonumber\\
    \xrightarrow{\hat{R}_y(\pi)}&\quad
        \left[\begin{array}{@{}cc@{}}
          \abs{\alpha}^2\left[\begin{matrix}
          E^1_{\pi}(\rho^4_1,\rho^1_1) & E^2_{\pi}(\rho^3_1,\rho^2_1) \\
          E^3_{\pi}(\rho^2_1,\rho^3_1) & E^4_{\pi}(\rho^1_1,\rho^4_1)
          \end{matrix}\right]
          & \alpha^*\beta\left[\begin{matrix}
          E^1_{\pi}(\rho^4_2,\rho^1_2) & E^2_{\pi}(\rho^3_2,\rho^2_2) \\
          E^3_{\pi}(\rho^2_2,\rho^3_2) & E^4_{\pi}(\rho^1_2,\rho^4_2)
          \end{matrix}\right] \\
          \alpha\beta^*\left[\begin{matrix}
          E^1_{\pi}(\rho^4_3,\rho^1_3) & E^2_{\pi}(\rho^3_3,\rho^2_3) \\
          E^3_{\pi}(\rho^2_3,\rho^3_3) & E^4_{\pi}(\rho^1_3,\rho^4_3)
          \end{matrix}\right] &
          \abs{\beta}^2\left[\begin{matrix}
          E^1_{\pi}(\rho^4_4,\rho^1_4) & E^2_{\pi}(\rho^3_4,\rho^2_4) \\
          E^3_{\pi}(\rho^2_4,\rho^3_4) & E^4_{\pi}(\rho^1_4,\rho^4_4)
          \end{matrix}\right]
        \end{array}\right]\nonumber\\
    \xrightarrow{\hat{S}_l}&\quad
        \left[\begin{array}{@{}cc@{}}
          \abs{\alpha}^2\left[\begin{matrix}
          \boxed{E^1_{\pi}(\rho^4_1,\rho^1_1)} & E^2_{\pi}(\rho^3_1,\rho^2_1) \\
          E^3_{\pi}(\rho^2_1,\rho^3_1) & \boxed{E^4_{\pi}(\rho^1_1,\rho^4_1)}
          \end{matrix}\right]
          & \alpha^*\beta\left[\begin{matrix}
          r^l_1 E^1_{\pi}(\rho^4_2,\rho^1_2) & \mathring{r}^l_1 E^2_{\pi}(\rho^3_2,\rho^2_2) \\
          \boxed{r^l_1 E^3_{\pi}(\rho^2_2,\rho^3_2)} & \mathring{r}^l_1 E^4_{\pi}(\rho^1_2,\rho^4_2)
          \end{matrix}\right] \\
          \alpha\beta^*\left[\begin{matrix}
          r^{l*}_1 E^1_{\pi}(\rho^4_3,\rho^1_3) & \boxed{r^{l*}_1 E^2_{\pi}(\rho^3_3,\rho^2_3)} \\
          \mathring{r}^{l*}_1 E^3_{\pi}(\rho^2_3,\rho^3_3) & \mathring{r}^{l*}_1 E^4_{\pi}(\rho^1_3,\rho^4_3)
          \end{matrix}\right] &
          \abs{\beta}^2\left[\begin{matrix}
          \boxed{\abs{r^l_1}^2 E^1_{\pi}(\rho^4_4,\rho^1_4)} & \mathring{r}^l_1 r^{l*}_1 E^2_{\pi}(\rho^3_4,\rho^2_4) \\
          \mathring{r}^{l*}_1 r^{l}_1 E^3_{\pi}(\rho^2_4,\rho^3_4) &   \boxed{\abs{\mathring{r}^l_1}^2 E^4_{\pi}(\rho^1_4,\rho^4_4)}
          \end{matrix}\right]
        \end{array}
        \right]\equiv\rho_{\text{out}}.\label{eq:spinflipfinal}
\end{align}
Here the basis states spanning $\rho_p\otimes\rho_s$ are $\{ \ket{e \Uparrow}_r,\ket{e \Downarrow}_r, \ket{l \Uparrow}_r, \ket{l \Downarrow}_r\}$ which govern only the Hilbert space formed by the reflected photon and spin, as events in which photons are transmitted do not contribute to the fidelity. Note that when applying the $\hat{R}_y(\pi)$ pulse, we apply Eq.~(\ref{eq:depolpi}) to each of the four $2\times2$ blocks (which consists of spin density matrix elements $\rho^i_j$). For instance, the inner product $\ket{e \Downarrow}_r\bra{e \Uparrow}_r$ has a matrix element $E^2_\pi (\rho^3_1,\rho^2_1)$, which corresponds to the $(1,2)$-th entry of the matrix $\mathbf{E}_\pi$ with $\rho_3\to\rho^3_1$ and $\rho_2\to\rho^2_1$. Only the boxed terms of Eq.~(\ref{eq:spinflipfinal}) contribute to the fidelity. As an example we evaluate one of the matrix elements $\ket{e \Uparrow}_r \bra{e \Uparrow}_r$:
\begin{align}
    \abs{\alpha}^2 E^1_\pi (\rho^4_1,\rho^1_1)&=\abs{\alpha}^2\bigg[(1-p_\pi)[\mathcal{F}_\pi \rho^4_1 + (1-\mathcal{F}_\pi)\rho^1_1] +\frac{p_\pi}{2}\bigg]=\abs{\alpha}^2\bigg[\frac{(1-p_\pi)}{2}[\mathcal{F}_\pi \abs{\mathring{r}^e_1}^2  + (1-\mathcal{F}_\pi)\abs{r^e_1}^2 ] +\frac{p_\pi}{2}\bigg].
\end{align}
For an ideal state of $\ket{\psi_{\text{ideal}}}=(\alpha\ket{e \Downarrow}_r - \beta\ket{l \Uparrow}_r)$ where $\abs{\alpha}=\abs{\beta}=1/\sqrt{2}$, the entanglement fidelity is given by
\begin{align}
        \mathcal{F}^{\text{theory}}_{\text{r}}
        \equiv\frac{\bra{\psi_{\text{ideal}}}\rho_{\text{out}}\ket{\psi_{\text{ideal}}}}{\Tr(\ket{\psi_{\text{out}}}\bra{\psi_{\text{out}}})}&=\frac{\abs{\alpha}^4 E^4_\pi(\rho^1_1,\rho^4_1) 
        + \abs{\beta}^4\abs{r^l_1}^{2}E^1_\pi(\rho^4_4,\rho^1_4)
        -\abs{\alpha^* \beta}^2 r^l_1 E^3_\pi(\rho^2_2,\rho^3_2)
        -\abs{\alpha \beta^*}^2 r^{l*}_1 E^2_\pi(\rho^3_3,\rho^2_3)
        }{\abs{\alpha}^2 E^4_\pi(\rho^1_1,\rho^4_1) 
        + \abs{\beta}^2\abs{r^l_1}^{2}E^1_\pi(\rho^4_4,\rho^1_4) + \abs{\alpha}^2 E^1_\pi(\rho^4_1,\rho^1_1) +\abs{\beta}^2 \abs{\mathring{r}^l_1}^{2}E^4_\pi(\rho^1_4,\rho^4_4) }\nonumber\\
        &\overset{\substack{r_1=-1 \\ \mathring{r}_1=0}}{\approx} 1-\frac{5\pi}{4}\bigg(\frac{\kappa+\gamma_r}{\Omega_r}\bigg)-\frac{3}{2}\frac{1}{\Omega^2_r {T^*_2}^2},\label{eq:sf}
\end{align}
for $\kappa\ll\Omega_r$ where $\Omega_r T_\pi=\pi$ for a $\pi$-pulse. The final expression is found by perturbative expansion for each error to the first order. Using the relevant parameters: $T_{\pi}=7$~ns, $T_{\pi/2}=3.5$~ns, $\kappa=0.0098\text{ ns}^{-1}$, $\gamma_r=0.0081\text{ ns}^{-1}$ and $T^*_2=23.2\text{ ns}$, we find $\mathcal{F}^{\text{theory}}_{\kappa}=84.6\%$ from the analytical form in Eq.~(\ref{eq:sf}) taking $r_1=-1$ and $\mathring{r}_1=0$.

\subsubsection{Spin readout error}
The non-ideal spin readout by optical pumping is also considered to be one of the dominant sources of imperfections as it directly influences the spin readout basis. Due to finite optical cyclicity, optically pumping of the main transition can unfavourably result in an opposite outcome by flipping the spin state:

\begin{align}
    \rho_{\text{out}}\quad\xrightarrow{\text{Spin readout}}\quad \mathcal{F}_{\text{R}}\rho_{\text{out}} + (1-\mathcal{F}_{\text{R}}) \hat{\sigma}_x(\pi) \rho_{\text{out}} \hat{\sigma}^\dagger_x(\pi).\label{eq:readoute}
\end{align}
where the readout fidelity is estimated to be $\mathcal{F}_{\text{R}}=96.6\%$~\cite{Appel2021thesis}. Using Eqs.~(\ref{eq:spinflipfinal}) and (\ref{eq:readoute}), the resulting entanglement fidelity under both rotation error and imperfect spin readout is $\mathcal{F}^{\text{theory}}_{\kappa,\text{R}}=(81.8\pm0.6)\%$. From here it is apparent that the dominant infidelity results from incoherent spin flips $\kappa$ and finite $\gamma_r$ ($15.4$\%). 

\subsubsection{Driving-induced dephasing due to multi-photon scattering}
Another source of error originates from the finite multi-photon component of the input pulse, which destroys the QD ground-state spin coherence through successions of photon-scattering events within the pulse. The driving-induced dephasing probability $p_d$ is related to the success probability of scattering $P_{\omega_1}+P_{\omega_2}$ and the mean photon number in the driving pulse $\bar{n}$ via $p_d=1-\exp[- \bar{n}(P_{\omega_1}+P_{\omega_2})]$~\cite{Elfving2019}. This can be understood as the probability of $\bar{n}$ disjoint successful scattering events. To describe the effect of this error, we adopt a phase-damping model $\mathcal{E}_d$ where
\begin{align}
    \mathcal{E}_d\bigg(\Tr_p\big(\hat{S}(\rho_p\otimes\rho_s)\big)\bigg) \equiv \bigg(1-\frac{p_d}{2}\bigg)\Tr_p\big(\hat{S}(\rho_p\otimes\rho_s)\big) + \frac{p_d}{2} \hat{\sigma}_z  \Tr_p\big(\hat{S}(\rho_p\otimes\rho_s)\big) \hat{\sigma}^\dagger_z=\left[\begin{matrix}
          s_{11}  & (1-p_d) s_{12}   \\
          (1-p_d) s_{21} & s_{22}
          \end{matrix}\right].
\end{align}
Here $\hat{S}$ is the scattering matrix acting on the spin-photon density matrix and $s_{ij}$ corresponds to the $(i,j)$-th entry of the reduced spin density matrix $\Tr_p\big(\hat{S}(\rho_p\otimes\rho_s)\big)$. $\mathcal{E}_d$ introduces dephasing only to the QD spin state thus the photonic component is traced out before applying the phase-damping channel. Now we follow the same approach in Sec.~\ref{subsec:spin-flip} and consider propagation of the dephasing error in the protocol:
\begin{align}
    \rho_p \otimes \rho_s
    \xrightarrow{\hat{R}_y(\pi/2)}& \left[\begin{array}{cc}
       \abs{\alpha}^2  &  \alpha^*\beta \\
        \alpha\beta^* & \abs{\beta}^2
    \end{array}\right] \otimes \frac{1}{2}\left[\begin{array}{cc}
       1  &  1 \\
        1 & 1
    \end{array}\right]\nonumber\\
    \xrightarrow{\mathcal{E}_d(\hat{S}_e)}&\quad
        \frac{1}{2}\left[\begin{array}{@{}cc@{}}
          \abs{\alpha}^2\mathcal{E}_d\left(\left[\begin{matrix}
          \abs{r^e_1}^2    &  r^{e*}_1 \mathring{r}^{e}_1  \\
          r^e_1 \mathring{r}^{e*}_1 & \abs{\mathring{r}^{e}_1}^2 
          \end{matrix}\right]\right)
          & \alpha^*\beta\mathcal{E}_d\left(\left[\begin{matrix}
          r^{e*}_1 &  r^{e*}_1 \\
           \mathring{r}^{e*}_1 & \mathring{r}^{e*}_1 
          \end{matrix}\right]\right) \\
          \alpha\beta^*\mathcal{E}_d\left(\left[\begin{matrix}
          r^e_1  & \mathring{r}^e_1  \\
          r^e_1  &\mathring{r}^e_1 
          \end{matrix}\right]\right) &
          \abs{\beta}^2\mathcal{E}_d\left(\left[\begin{matrix}
          1 & 1 \\
          1 & 1
          \end{matrix}\right]\right)
        \end{array}
        \right]\nonumber\\
    \xrightarrow{\hat{R}_y(\pi)}&\quad
        \frac{1}{2}\left[\begin{array}{@{}cc@{}}
          \abs{\alpha}^2\left[\begin{matrix}
          \abs{\mathring{r}^{e}_1}^2    &  -(1-p_d) r^e_1 \mathring{r}^{e*}_1  \\
          -(1-p_d) r^{e*}_1 \mathring{r}^{e}_1 & \abs{r^e_1}^2 
          \end{matrix}\right]
          & \alpha^*\beta\left[\begin{matrix}
          \mathring{r}^{e*}_1 &  -(1-p_d)\mathring{r}^{e*}_1 \\
           -(1-p_d) r^{e*}_1 & r^{e*}_1 
          \end{matrix}\right] \\
          \alpha\beta^*\left[\begin{matrix}
          \mathring{r}^e_1  & -(1-p_d) r^e_1  \\
          -(1-p_d)\mathring{r}^e_1  &r^e_1 
          \end{matrix}\right] &
          \abs{\beta}^2\left[\begin{matrix}
          1 & -(1-p_d) \\
          -(1-p_d) & 1
          \end{matrix}\right]
        \end{array}
        \right]\nonumber\\
    \xrightarrow{\mathcal{E}_d(\hat{S}_l)}&\quad
        \frac{1}{2}\left[\begin{array}{@{}cc@{}}
          \abs{\alpha}^2\left[\begin{matrix}
          \abs{\mathring{r}^{e}_1}^2    &  -(1-p_d)^2 r^e_1 \mathring{r}^{e*}_1  \\
          -(1-p_d)^2 r^{e*}_1 \mathring{r}^{e}_1 & \abs{r^e_1}^2 
          \end{matrix}\right]
          & \alpha^*\beta\left[\begin{matrix}
          r^l_1 \mathring{r}^{e*}_1    &  -(1-p_d)^2 \mathring{r}^l_1 \mathring{r}^{e*}_1   \\
          -(1-p_d)^2 r^l_1 r^{e*}_1  & \mathring{r}^l_1 r^{e*}_1 
          \end{matrix}\right] \\
          \alpha\beta^*\left[\begin{matrix}
          r^{l*}_1 \mathring{r}^e_1   &  -(1-p_d)^2 r^{l*}_1 r^e_1   \\
          -(1-p_d)^2 \mathring{r}^{l*}_1 \mathring{r}^e_1  & \mathring{r}^{l*}_1 r^e_1
          \end{matrix}\right] &
          \abs{\beta}^2\left[\begin{matrix}
              \abs{r^l_1}^2     &  -(1-p_d)^2 r^{l*}_1 \mathring{r}^l_1   \\
          -(1-p_d)^2 r^{l}_1 \mathring{r}^{l*}_1  & \abs{\mathring{r}^l_1}^2  
          \end{matrix}\right]
        \end{array}
        \right].
\end{align}
Similarly, the entanglement fidelity under the driving-induced dephasing is found to be
\begin{align}
    \mathcal{F}^{\text{theory}}_{\text{r}} = \frac{\frac{1}{2}\abs{r_1}^2\big[1+e^{-2\bar{n}(P_{\omega_1}+P_{\omega_2})}\big]}{\abs{r_1}^2+\abs{\mathring{r}_1}^2}\xrightarrow{r_1\to-1,\text{ }\mathring{r}_1\to0}\frac{1}{2}\bigg[1+e^{-2\bar{n}(P_{\omega_1}+P_{\omega_2})}\bigg]\approx1-\bar{n}(P_{\omega_1}+P_{\omega_2}),\label{eq:drivingd}
\end{align}
where the average probability of successful scattering is given by Eq.~(\ref{eq:scf}):
\begin{align}
    P_{\omega_1}+P_{\omega_2}=\int^{\infty}_{-\infty}\bigg[\abs{r_1(\omega)}^2+\abs{t_1(\omega)}^2+\abs{r_2(\omega)}^2+\abs{t_2(\omega)}^2\bigg]\abs{\Phi_1(\omega)}^2d\omega.
\end{align}
To estimate the infidelity in the experiment, we first extract the average number of photons in the pulse $\bar{n} \leq (0.089\pm0.012)$ (Supplementary Note 5). Given that optical cyclicity $C=14.7$, pulse bandwidth $\sigma_o=\sqrt{2\ln{2}}/T_{\text{FWHM}}\approx0.589\text{ ns}^{-1}$, spectral diffusion fluctuation $\sigma_e=2\pi\times(332\pm15)$~MHz and waveguide coupling efficiency $\beta\geq0.865\pm0.059$ (Supplementary Note 4), the experimental infidelity is estimated using the exact form in Eq.~(\ref{eq:drivingd}) to be $1-\mathcal{F}^{\text{theory}}_{\bar{n}}\leq(7.2\pm0.7)\%$.
\subsection{Estimate the Overall Entanglement Fidelity}
\label{subsec:overall}
Assuming perfect manipulation of the hole spin state, the entanglement fidelity is expressed by:
\begin{align}
    \mathcal{F}^{\text{theory}}_{\text{r}}=1-\frac{\gamma_d}{\Gamma}-\frac{\Gamma^2}{4\Delta^2_h}\bigg(\frac{C}{C+1}\beta\bigg)^2\approx1-\frac{\gamma_d}{\Gamma}-\frac{\Gamma^2}{4\Delta^2_h},
\end{align}
which is estimated to be
$(96.2\pm0.1)\%$ with $\Gamma=2.48 \text{ ns}^{-1}$, $\Delta_h=2\pi\times7.3$~GHz~\cite{Appel2021} and $\gamma_d=(0.099\pm0.004)\text{ ns}^{-1}$ (Supplementary Note 3). This predominantly reflects the infidelity from phonon-induced pure dephasing $1-\mathcal{F}^{\text{theory}}_{\gamma_d}$ as the off-resonant reflection error $\Gamma^2/\Delta^2_h$ is comparably small. Together with the incoherent spin-flip, driving-induced dephasing and the readout errors discussed above, we estimate a lower bound on the overall entanglement fidelity $\mathcal{F}^{\text{theory}}_{\text{total}}$ of
\begin{align}
    \mathcal{F}^{\text{theory}}_{\text{total}} \approx \mathcal{F}^{\text{theory}}_{\gamma_d} \times \mathcal{F}^{\text{theory}}_{\kappa,\text{R}}\times \mathcal{F}^{\text{theory}}_{\bar{n}} \approx (73.0\pm0.6)\%,
\end{align}
which generally agrees with the experimentally obtained value $(74.3\pm2.3)\%$ including error margins. Here a lower fidelity bound is obtained as the waveguide-coupling efficiency $\beta$ could be underestimated (Supplementary Note 4) which leads to overestimating $\bar{n}$ and the corresponding infidelity $1-\mathcal{F}^{\text{theory}}_{\bar{n}}$. Imperfect spin initialization ($1.4\%$)~\cite{Appel2021b} is not considered in the theory but is expected to have negligible infidelity ($<1\%$).
\newpage
\section*{Supplementary Note 3: Photon Visibility and Pure Dephasing Rate Estimation}
\label{sec:vis}
Here we derive an analytical form of the visibility as a function of the QD pure dephasing rate. In the experiment, a time-bin encoded qubit (a weak coherent state) is scattered by a QD spin embedded in a two-sided photonic-crystal waveguide, and is subsequently measured by an asymmetric Mach-Zehnder interferometer with equal time delay as the qubit. The visibility is therefore a measure of the temporal overlap between the time-bins of the scattered pulses. To model this, we consider the scattering of the time-bin photon with the QD and project the output state onto the photonic X-bases. The initial state of the system is expressed as $\ket{\text{in}}=(\ket{e}+ \ket{l})/\sqrt{2} \otimes\ket{\Uparrow}$. Here we have neglected the multi-photon components from the coherent state since we are interested in the effect of pure dephasing. For a complete modelling of the photon visibility, however, one should include the effect of multi-photon scattering and inelastic contributions~\cite{Shen2007}.

With Eq.~(\ref{eq:scf}) the output state becomes
\begin{align}
    \ket{\text{out}} &=  \frac{1}{\sqrt{2}}\bigg[r_1 \bigg(   \ket{e}_r + \ket{l}_r \bigg) + t_1 \bigg(   \ket{e}_t + \ket{l}_t \bigg) \bigg] \otimes \ket{\Uparrow} + \frac{1}{\sqrt{2}}\bigg[r_2 \bigg(   \ket{e'}_r + \ket{l'}_r \bigg)+  t_2 \bigg(   \ket{e'}_t + \ket{l'}_t \bigg)\bigg]\otimes\ket{\Downarrow},
\end{align}
where the superscript prime ($'$) represents a scattered photon of frequency $\omega_2\neq\omega_1$ and the subscript ``r" (``t") indicates a reflected (transmitted) photon.
 We then seek the photonic density matrix by tracing out the spin degree of freedom, the transmitted photons as well as the wrong frequency state $\omega_2$. For ease of computation the scattering coefficients are replaced by $C_{i}$ where $i$ refers to the time-bin, thus
\begin{align}
    \ket{\text{out}}_{\text{p}}\bra{\text{out}}_{\text{p}} &= \frac{\Tr_{s,t,\omega}(\ket{\text{out}}\bra{\text{out}})}{\Tr(\ket{\text{out}}\bra{\text{out}})} =\abs{C_{e}}^2 \ket{e}\bra{e} + \abs{C_{l}}^2 \ket{l}\bra{l} +C_{e} C_{l}^* \ket{e}\bra{l}+C_{l} C_{e}^* \ket{l}\bra{e}.\label{eq:outputp}
\end{align}
Now Eq.~(\ref{eq:outputp}) is used to evaluate the middle-bin intensity in detector D2(D1):
\begin{align}
    \text{I}_{\text{D2/D1}} = \int\Tr\bigg[ \frac{(\hat{a}_e \pm e^{i\theta_{p}} \hat{a}_l)}{\sqrt{2}} \bigg(\ket{\text{out}}_{\text{p}}\bra{\text{out}}_{\text{p}}\bigg) \frac{(\hat{a}^\dagger_e \pm e^{-i\theta_{p}} \hat{a}^\dagger_l)}{\sqrt{2}} \bigg]dt,
\end{align}
where the output photon state is projected onto the superposition state $\hat{a}_e(t) \pm e^{i\theta_{p}} \hat{a}_l(t)$ which is equivalent to adding a phase shifter on the long path of the excitation interferometer and interfering both bins. Setting $\theta_p=0$ implies projecting the output state into the $p_\pm = \ket{\pm X}_p\bra{\pm X}_p$ bases as described in the main text. The projected state is then traced out in both the early and late time bases. The photon visibility is defined as the normalized contrast of the middle-bin intensity when $\theta_p=0$:
\begin{align}
    V_p \equiv \frac{\text{I}_{\text{D2}}-\text{I}_{\text{D1}}}{\text{I}_{\text{D2}}+\text{I}_{\text{D1}}}=\frac{ \int \Tr\bigg( \hat{a}_e(\ket{\text{out}}_{\text{p}}\bra{\text{out}}_{\text{p}})\hat{a}^\dagger_l + \hat{a}_l(\ket{\text{out}}_{\text{p}}\bra{\text{out}}_{\text{p}})\hat{a}^\dagger_e  \bigg) dt }{\int \Tr\bigg( \hat{a}_e(\ket{\text{out}}_{\text{p}}\bra{\text{out}}_{\text{p}})\hat{a}^\dagger_e + \hat{a}_l(\ket{\text{out}}_{\text{p}}\bra{\text{out}}_{\text{p}})\hat{a}^\dagger_l  \bigg) dt}.\label{eq:v}
\end{align}
To further simplify the above expression, we consider the scattering events of the early and late bins to be identical, i.e., with the same scattering coefficient $C_{e}=C_{l}=r_1$. as justified in Sec.~\ref{subsubsec:spind}. Therefore, under this assumption the photon visibility becomes unity in the single-photon regime.

Following from the discussion in Sec.~\ref{subsubsec:pured}, we can now account for the effect of phonon-induced pure dephasing~\cite{Tighineanu2018}. In essence, the resulting spin-photon density matrix is the sum of coherent and incoherent parts as described by Eqs.~(\ref{eq:1phonon}) and (\ref{eq:pured}). The advantage of the formalism in Eq.~({\ref{eq:1phonon}}) is that its effect can be straightforwardly included in Eq.~(\ref{eq:outputp}). Accordingly, the new photonic density matrix becomes
\begin{align}
    \ket{\text{out}'}_{\text{p}}\bra{\text{out}'}_{\text{p}} &\approx \ket{\text{out}}_{\text{p}}\bra{\text{out}}_{\text{p}} +  \frac{1}{2} P^{\omega_1}_{\gamma_d}\rho^{\omega_1}_{\gamma_d,\text{e}} \ket{\emptyset}_l \bra{\emptyset}_l +  \frac{1}{2} P^{\omega_1}_{\gamma_d}\rho^{\omega_1}_{\gamma_d,\text{l}} \ket{\emptyset}_e \bra{\emptyset}_e,
\end{align}
where the last two terms correspond to dephasing occurred during the single-photon scattering of either the early or late time-bin. The effect of pure dephasing on the multi-photon component is not considered due to its polynomial dependence on the mean photon number per pulse $\bar{n}$, which is negligible as $\bar{n}\ll1$. Note that the incoherent photon does not contribute to the interference since $\Tr(\hat{a}_e \rho^{\omega_1}_{\gamma_d,\text{e}} \ket{\emptyset}_l \bra{\emptyset}_l \hat{a}^\dagger_l) = \Tr(\hat{a}_e \rho^{\omega_1}_{\gamma_d,\text{e}} ) \times \Tr(\ket{\emptyset}_l \bra{\emptyset}_l \hat{a}^\dagger_l)=0$. This means only the total intensity is affected and Eq.~(\ref{eq:v}) can be simplified as
\begin{align}
    V_p = \frac{\int^{\infty}_{-\infty}\abs{\Phi_1(\omega)}^2\abs{r_1(\omega)}^2 d\omega }{ \int^{\infty}_{-\infty}\abs{\Phi_1(\omega)}^2\abs{r_1(\omega)}^2 d\omega+  P^{\omega_1}_{\gamma_d} }.\label{eq:Vpd}
\end{align}
To include slow resonance drift $\omega\to\omega+\delta_e$ due to spectral wandering, both the probability of reflecting a photon $\int^{\infty}_{-\infty}\abs{\Phi_1(\omega)}^2\abs{r_1(\omega)}^2 d\omega$ and the pure dephasing probability $P^{\omega_1}_{\gamma_d}$ need to be averaged over a Gaussian distribution $N(0,\sigma_e)$:
\begin{align}
    N(0,\sigma_e)=\frac{1}{\sqrt{2\pi\sigma^2_e}}e^{-\frac{\delta_e^2}{2\sigma^2_e}},\label{eq:sdGauss}
\end{align} with standard deviation of the drift $\sigma_e$~\cite{LeJeannic2021}. Eq.~(\ref{eq:pured}) is then rewritten as
\begin{align}
    P^{\omega_1}_{\gamma_d} = \frac{\Gamma^r_1}{\Gamma} P_{\gamma_d}&= \frac{\Gamma^r_1}{\Gamma}\int^{\infty}_{-\infty}\int^{\infty}_{-\infty} \frac{e^{-\frac{(\omega-\omega_1)^2}{2\sigma^2_o}}}{\sqrt{2\pi\sigma^2_o}}\abs{\frac{-2\sqrt{2\gamma_d\Gamma^r_1}}{\Gamma+2i(\omega-\omega_1+\delta_e)}}^2 N(0,\sigma_e)d\omega d\delta_e,\label{eq:puredspec}
\end{align}
and Eq.~(\ref{eq:Vpd}) becomes
\begin{align}
    V_p = \frac{\int^{\infty}_{-\infty}\int^{\infty}_{-\infty}\abs{\Phi_1(\omega)}^2\abs{r_1(\omega)}^2 N(0,\sigma_e) d\omega d\delta_e }{ \int^{\infty}_{-\infty}\int^{\infty}_{-\infty}\abs{\Phi_1(\omega)}^2\abs{r_1(\omega)}^2 N(0,\sigma_e) d\omega d\delta_e+  P^{\omega_1}_{\gamma_d} }=\frac{\Gamma}{\Gamma+2\gamma_d},\label{eq:Vpd2}
\end{align}
which is interestingly identical to the photon indistinguishability formula derived in Ref.~\cite{Tighineanu2018}. In the limit $\bar{n}\approx0$ where only single photons interact with the QD, the $y$-intercept of the photon visibility curve in Figure~2b of the main text is therefore given by Eq.~(\ref{eq:Vpd2}). A linear fit followed by extrapolation of the data gives a $y$-intercept of $V_p=0.926\pm0.003$ implying $\gamma_d\approx(0.099\pm0.004)\text{ ns}^{-1}$. $V_p$ can be intuitively understood as the mean wavepacket overlap between single photons.

\section*{Supplementary Note 4: Two-color Transmission Experiment}
\label{sec:RT}
\begin{figure}[h]
     \centering
     \includegraphics[scale=0.9]{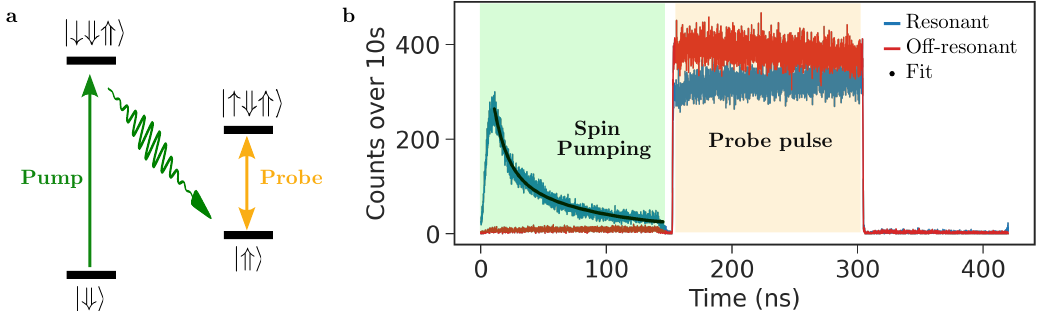}
     \caption{\textbf{Two-color transmission measurement on the waveguide-embedded QD hole spin.} (a) QD Level diagram. A strong pump pulse (green) prepares the spin state $\ketUp$, while the frequency of the weak probe pulse (orange) is scanned to reveal the QD transmission spectrum. (b) Time-resolved histogram of the pulse sequence with resonant (1.148~V) and off-resonant (1~V) bias voltages. A bi-exponential fit is used to extract a lower bound on the spin pumping fidelity. }\label{fig:RThist}
\end{figure}
To probe the photon-emitter coupling efficiency of the device and spectral diffusion noise of the QD spin, we perform a series of spin-dependent transmission measurements at various powers of the probe laser. The experiment is conducted by first initializing the QD spin in the $\ket{\Uparrow}$ state, followed by scanning the frequency of a 150-ns probe pulse at a given power~(Supplementary Figure~\ref{fig:RThist}). Since the probe is orders of magnitude longer than the QD lifetime, the atom saturates similarly as being driven by a continuous-wave laser. The destructive interference between the phase-shifted scattered light and the probe in the waveguide results in a transmission dip on resonance~\cite{Thyrrestrup2018,LeJeannic2021}, where the shape of the transmission spectrum is determined by the combination of phonon-induced pure dephasing $\gamma_d$, coupling efficiency $\beta$, spectral diffusion $\sigma_e$, Fano parameter $\xi$, laser power $P$ and cyclicity $C$. In particular, the normalized transmission function used to fit the data is given by
\begin{align}
    \mathcal{I}_T(\omega-\omega_1)  = 1+P_{sp}\times \Re\bigg(\frac{\frac{C}{C+1}\beta \Gamma^2 \big[\Gamma+2\gamma_d+2i(\omega-\omega_1)\big](i+\xi)\big[2i+\frac{C}{C+1}\beta(-i+\xi)\big]}{\Gamma\big[(\Gamma+2\gamma_d)^2+4(\omega-\omega_1)^2\big]+8\eta (\Gamma+2\gamma_d)P}\bigg).\label{eq:RT2}
\end{align}
Here $P_{sp}$ represents the spin pumping fidelity. A non-unity $P_{sp}$ implies the QD could go dark when its spin is prepared in the wrong state~\cite{Javadi2018}. An average value for $P_{sp}$ is obtained by first fitting the fluorescence decay during the pumping pulse (Supplementary Figure~\ref{fig:RThist}b) then averaging over all frequencies and probe powers. In the low power limit $P\to0$ with a perfectly prepared two-level system $P_{sp}\to1$, the expression reduces to Eq.~(S33) in Ref.~\cite{LeJeannic2021} with $\beta$ replaced by $\frac{C}{C+1}\beta$, as a finite $C$ limits the number of resonantly reflected photons leading to diminished interference. To take slow resonance drifts due to spectral diffusion into account, $\mathcal{I}_T(\omega-\omega_1)$ is averaged over the Gaussian distribution $N(0,\sigma_e)$ (Supplementary Note 3). 

\begin{figure*}[ht]
     \centering
     \includegraphics[scale=0.9]{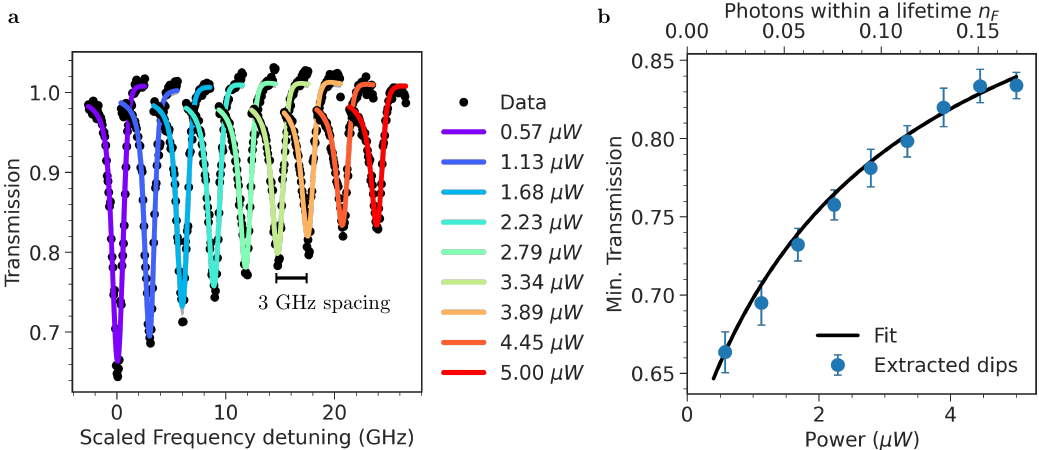} 
     \caption{\textbf{Results from iteratively fitting transmission spectra to extract coupling efficiency $\beta$ and spectral diffusion $\sigma_e$.} (a) Transmission spectra fitted at different probe powers. The frequency detuning axis has been rescaled such that each spectra is $3$~GHz apart. The central frequency is 317.235~THz. The data (black circles) are fitted using Eqs.~(\ref{eq:sdGauss}) and  (\ref{eq:RT2}). (b) The transmission dips extracted from (a) are then fitted to estimate $\beta$.}\label{fig:RT}
\end{figure*}

We have developed an iterative fitting procedure for the data in Supplementary Figure~\ref{fig:RT} to reliably estimate $\beta$ and $\sigma_e$. The general idea is to divide the data into two sub-dataset and fit both on each iteration. The first dataset is a set of transmission spectra as a function of probe powers, while the second dataset consists of only the transmission dip at various powers. The algorithm runs iteratively based on results from the previous fit and terminates when the fitted parameters from both fits converge. The convergence implies that there is a set of parameters which simultaneously holds true when using two different fit functions on the same data.

The fitting process is described as follows: (1) We fix the total decay rate $\Gamma=2.48\text{ ns}^{-1}$~\cite{Appel2021b}, $P_{sp}=0.897$, $\gamma_d=0.099\text{ ns}^{-1}$ and $C=14.7$~\cite{Appel2021b} as they are measured independently. For the first fit we assume $\beta=0.95$; (2) Based on these values, we take $\sigma_e$, $\xi$ and loss factor $\eta$ as free parameters to perform a least square fit on the transmission spectrum at each power (Supplementary Figure~\ref{fig:RT}a). This results in a list of fitted values for $\sigma_e$, $\xi$ and $\eta$. Their corresponding mean values are then used to fit the second dataset (transmission dips as a function of probe power) with only $\beta$ and $\eta$ as free parameters (Supplementary Figure~\ref{fig:RT}b). (3) From here we obtain an updated value of $\beta$ which is used to fit the transmission spectra again in step 2. (4) The iteration stops when $\beta$ after loop $i$ converges (i.e., $\abs{\beta_i - \beta_{i-1}}<0.1\%$). 

\begin{table}[ht]
    \centering
    \begin{tabular}{c c c}
    \toprule
    \textbf{Parameter} & \textbf{Value} & \textbf{Confidence interval (99.7\%)} \\ 
    \colrule
    $\beta$& 0.865 & [0.806, 0.924]\\
    $\sigma_{e}/2\pi$& 332 MHz & [317, 347]\\
    $\eta$& 0.427 & [0.406, 0.448]\\
    $\xi$& -0.127 & [-0.136, -0.118]\\
    $P_{sp}$& 0.897 & [0.895, 0.899]\\
    \botrule
    \end{tabular}
    \caption{Relevant parameters extracted from fitting the transmission spectra.}
    \label{table:RT}
\end{table}
The fit is completed in 10 iterations. The extracted parameters with $3\sigma$-uncertainty are presented in Supplementary Table~\ref{table:RT}. Both $\sigma_e=2\pi\times(332\pm15)$~MHz and $\beta=(0.865\pm0.059)$ are in very good agreement with previous estimates from two-color continuous-wave pumping resonance fluorescence~\cite{Appel2021} and transmission~\cite{LeJeannic2021} measurements, respectively, indicating that the combination of two-color pulsed transmission and photon visibility measurements through QD scattering in the waveguide could be an alternative way to accurately extract these QD noise parameters. Note that due to the non-unity hole initialization efficiency, the actual value of $\beta$ could be even higher, as the QD also blinks if the hole spin is not loaded. Additionally, since the pumping pulse is generated with a slow acousto-optical modulator with 8~ns rise time, the imperfect pump pulse shape together with residual repumping from the probe pulse might have underestimated the spin pumping fidelity. Therefore, the extracted value of $\beta=(0.865\pm0.059)$ constitutes a lower bound.

The conversion between the probe power and the mean photon number \textit{within} a QD lifetime $n_F$ in Supplementary Figure~\ref{fig:RT}b is obtained using Eq.~(\ref{eq:saturation}) and the extracted values for $\eta$, $\sigma_e$, $\beta$ (Supplementary Table~\ref{table:RT}) and $\gamma_d$. The transmission spectrum in Figure~1c (in the main text) is probed at an input power of $P=0.57~\mu$W, with the saturation parameter $S=0.054$ corresponding to an average of $n_F=0.02\ll 1$ photons interacting with the QD within its lifetime.
\section*{Supplementary Note 5: Spin-dependent Reflectivity Measurement}
\label{sec:photonnumber}

\begin{figure}[ht]
     \centering
     \includegraphics[scale=0.9]{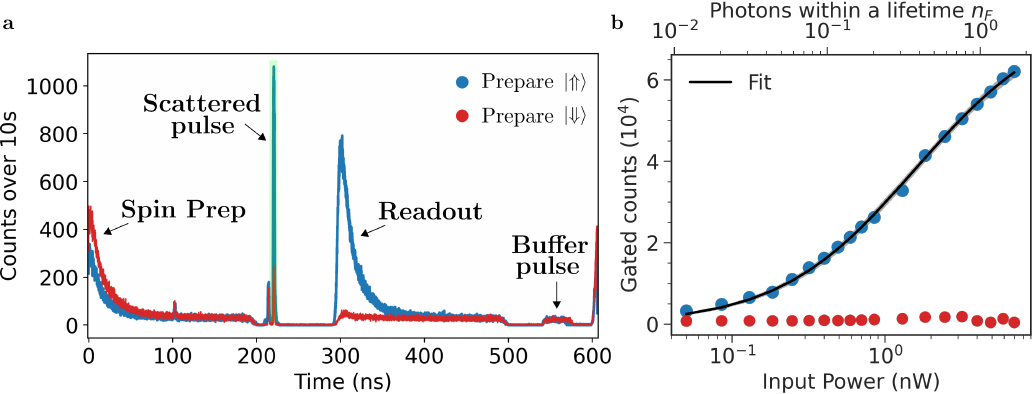}
     \caption{\textbf{Saturation measurement to calibrate the mean photon flux.} \textbf{(a)} Time-resolved histogram of the measurement sequence. A $2$ ns pulse gets reflected from a QD prepared in $\ket{\Uparrow}$ via optical spin pumping followed by a $\pi$-rotation pulse. The reflected signal is time-gated (green shaded region) and recorded for each input power. Peaks at around $100$ and 215~ns are laser scatter from the time-bin interferometer and the optical breadboard respectively. The spin readout at 300~ns maintains the same duty cycle as the entanglement experiment and does not affect the gated counts. \textbf{(b)} Gated fluorescence in the reflection as a function of the input pulse power. Blue (red) circles are summed counts over a time window of 3~ns, when the QD spin is prepared in $\ket{\Uparrow}$ ($\ket{\Downarrow}$). Fitted (black solid line) using Eq.~(\ref{eq:saturation}). Around 0.075~nW is used for a single pulse in the entanglement experiment.}
     \label{fig:saturation}
\end{figure}

Apart from measuring the photon visibility, another approach to probe the single-photon nature of the scattering process is through QD saturation measurement, in which the QD response is observed by scanning the power of the input qubit laser. From fitting the scattered signal, the mean photon number per pulse $\bar{n}$ can be extracted, where $\bar{n}\ll1$ indicates the scattering occurs in the single-photon regime. To mimic the entanglement experiment, we prepare a single pulse of 2~ns  duration and scatter on a QD spin initialized in either $\ket{\Uparrow}$ or $\ket{\Downarrow}$. Due to the QD spin-dependent reflectivity, the input photon which is resonant with the QD transition $\ket{\Uparrow}\to\ketupe$ is coherently reflected. By time-gating on the reflected signal (Supplementary Figure~\ref{fig:saturation}a; green shaded region) and increasing the input power, the QD prepared in $\ket{\Uparrow}$ becomes saturated (Supplementary Figure~\ref{fig:saturation}b). The averaged intensity in the reflected signal is fitted assuming a two-level system between $\ket{\Uparrow}\to\ketupe$:
\begin{align}
    \mathcal{I}_R = \mathcal{I}_{\text{max}} \int^{\infty}_{-\infty}\frac{\frac{C}{C+1}\beta(1+\frac{2\gamma_d}{\Gamma})\Omega^2_1}{(\frac{\Gamma}{2}+\gamma_d)^2+\delta^2_e+2(1+\frac{2\gamma_d}{\Gamma})\Omega^2_1}N(0,\sigma_e)d\delta_e,\label{eq:saturation}
\end{align}
where $\Omega_1$ is the Rabi frequency driving the transition $\ket{\Uparrow}\to\ketupe$, $\delta_e$ is the effective resonance drift due to spectral diffusion. A setup loss factor $\eta$ is introduced to associate the Rabi frequency to the input power $P$ where $\Omega_1= \sqrt{\eta P}$. Here $\mathcal{I}_{\text{max}}$ and $\eta$ are free parameters, whereas $\sigma_e$ and $\gamma_d$ are fitted in Supplementary Note 4.
Eq.~(\ref{eq:saturation}) holds when $\Gamma\gg\kappa_g$ and $T_p\gg\Gamma^{-1}$ where $\kappa_g$ is the effective spin-flip rate between the hole ground states and $T_p$ is the FWHM qubit duration measured from the pulse intensity. The first condition implies that the main transition is eventually saturated as the QD decays faster than the spin can recycle, thus $\ket{\Downarrow}$ effectively becomes dark. This is generally true since $\kappa_g$ is typically on the order of $10^{-7}\text{ ns}^{-1}$ at the plateau center voltage~\cite{Dreiser2008}, which is lower than $\Gamma=2.48\text{ ns}^{-1}$. The second condition ensures that the QD decays back to $\ket{\Uparrow}$ before the next scattering event within the pulse. When the driving pulse is sufficiently long, i.e., $T_p=2\text{ ns}>\Gamma^{-1}=0.4\text{ ns}$ with increasing power, the QD saturates similarly as when being driven by a weak continuous-wave laser. In addition, a finite cyclicity leads to a resonant spin-flip into the dark state $\ket{\Downarrow}$ reflecting a photon of frequency $\omega_2\neq\omega_1$ which is filtered out, thus only reducing the total intensity included in $\mathcal{I}_{\text{max}}$ and not affecting the scaling of Eq.~(\ref{eq:saturation}).

From fitting the data (Supplementary Figure~\ref{fig:saturation}b), we extract $\eta=(3.00\pm0.07)\text{ ns}^{-2}$/nW. The saturation parameter $S$ at an input power $P=0.075$~nW used for a single pulse is estimated to be $0.05\pm0.001$ using
\begin{align}
    S = \int^{\infty}_{-\infty} \frac{2(1+\frac{2\gamma_d}{\Gamma}) \eta P}{(\frac{\Gamma}{2}+\gamma_d)^2+\delta^2_e} N(0,\sigma_e)d\delta_e.
\end{align}
The mean photon number \textit{within} QD lifetime or mean photon flux is defined as $n_F\equiv S n_c$ with $n_c\equiv(1+2\gamma_d/\Gamma)/4\beta^2$ to be the critical photon flux leading to an excited state population of $1/4$. As a sanity check, in the ideal limit where $\gamma_d,\delta_e \to0$ and $\beta\to1$, $n_F=2\Omega^2_1/\Gamma^2$ which recovers the definition in Ref.~\cite{Thyrrestrup2018}. Using $\beta\geq0.865\pm0.059$, we estimate the average number of photons in a single pulse $\bar{n}=n_F T_p \Gamma\leq0.089\pm0.012\ll1$.
\newpage
\section*{Supplementary Note 6: Optical Spin Control and Rotation Fidelity}
\label{sec:rotation}
In this section, we briefly summarize the principle of optical control of the hole spin ground states, followed by experimental characterization of Rabi flopping between the hole ground states to extract the $\pi$-rotation fidelity.
\subsection*{Principle of Optical Spin Control}
Supplementary Figure~\ref{fig:Rabi_fit}b depicts the level scheme of a hole spin under an external in-plane magnetic field, under the driving of a bi-chromatic laser. To optically drive the spin transitions, we employ a monochromatic laser (green) at a frequency $\omega_o=\omega_1-\Delta_r$ detuned from the main transition $\ketUp\leftrightarrow\ketupe$. This laser is microwave-modulated resulting in two sidebands at $\omega_o\pm\Delta_h/2$, where the higher-frequency (lower-frequency) band drives the $\ketDown$ ($\ketUp$) state. The combination of both colors drives both $\Lambda$-systems (labelled as ``1" and ``2") via two-photon Raman transitions without populating the trions, thus creating an effective coupling between only the hole ground-state manifold.

For this two-photon Raman scheme to work~\cite{Bodey2019,Appel2021b}, each frequency band must be circularly-polarized, such that (1) it drives both vertical Y-polarized and diagonal X-polarized transitions with equal optical Rabi frequencies ($\Omega_y=\Omega_x$), and (2) $\Omega_x$ is $\pi/2$ out of phase with $\Omega_y$, thus the optical fields constructively drive both $\Lambda$-systems. Intuitively, this could be visualized as reversing the direction of the arrows for diagonal transitions in Supplementary Figure~\ref{fig:Rabi_fit}b, where the $\Lambda$-system ``1" (``2") now transfers population from $\ketDown$ to $\ketUp$ ($\ketUp$ to $\ketDown$), forming a cycling drive via a virtual state.

Mathematically, the spin dynamics under the optical drive is governed by the effective spin TLS Hamiltonian~\cite{Appel2021thesis}:
\begin{align}
    \hat{H}_{s} = \frac{\Omega_r}{2}(\hat{\sigma}_x\cos\phi_r  + \hat{\sigma}_y\sin\phi_r ) - \frac{\Delta_{\text{MW}}}{2}\hat{\sigma}_z,\label{eq:spinHamil}
\end{align}
where $\Omega_r$ is the effective spin Rabi frequency of the TLS. $\phi_r$ is the azimuthal angle on the Bloch sphere with hole spin states as the poles. $\phi_r=0$ ($\phi_r=\pi/2$) sets the axis of rotation along $+x$ ($+y$). $\hat{\sigma}_i$ are Pauli matrices. $\Delta_{\text{MW}}$ is the detuning of the bi-chromatic laser with the virtual state. To study the time evolution of the hole spin state under Eq.~(\ref{eq:spinHamil}), we refer to the Lindblad master equation ($\hbar=1$):
\begin{align}
    \dot{\rho}_s = -i [ \hat{H}_s,\rho_s ] + \sum^n_{i=1} D[\hat{C}_i]\rho_s, \label{eq:spinTLSH}
\end{align}
for $\rho_s$ is the spin density matrix spanning the hole spin ground states. The latter term in Eq.~(\ref{eq:spinTLSH}) represents the static Lindblad dissipator $D[\hat{C}_i]\rho\equiv\hat{C}_i\rho\hat{C}_i^\dagger-\frac{1}{2}\{\hat{C}_i^\dagger\hat{C}_i,\rho\}$ for a list of $n$ collapse operators $\hat{C}_i$. To model realistic noises in the experiment, we first consider a Markovian noise that destroys spin coherence by flipping the spin state with a rate $\kappa$. Specifically, we take $\hat{C}_1=\sqrt{\kappa}\hat{\sigma}_{+}$ and $\hat{C}_2=\sqrt{\kappa}\hat{\sigma}_{-}$ where $\hat{\sigma}_{+}\equiv\ketDown\bra{\Uparrow}=\hat{\sigma}^\dagger_{-}$ is the atomic raising operator.

Generally, for a finite detuning $\Delta_{\text{MW}}\neq 0$, Eq.~(\ref{eq:spinTLSH}) can only be numerically solved. To elaborate further, we now include the Overhauser field noise. For a frozen nuclear bath~\cite{Merkulov2002}, this corresponds to introducing a fluctuating noise $\Delta_{\text{MW}}=\delta_g$ along $z$-axis of the spin Bloch sphere. The $\ketDown$ population under such noise is then found by averaging $\rho_{\ketUp\bra{\Uparrow}}(T_r,\delta_g)$ with a Gaussian spin dephasing profile $N(\delta_g,\sigma_{\text{OH}})$:
\begin{align}
    \bar{\rho}_{\ketUp\bra{\Uparrow}}(T_r)=\int^{\infty}_{-\infty}d\delta_g~\rho_{\ketUp\bra{\Uparrow}}(T_r,\delta_g)~N(\delta_g,\sigma_{\text{OH}}).\label{eq:Rabifit}
\end{align}
\subsection*{Measured Spin Rabi Oscillations}
\begin{figure}[ht]
    \centering
	\includegraphics[scale=0.9]{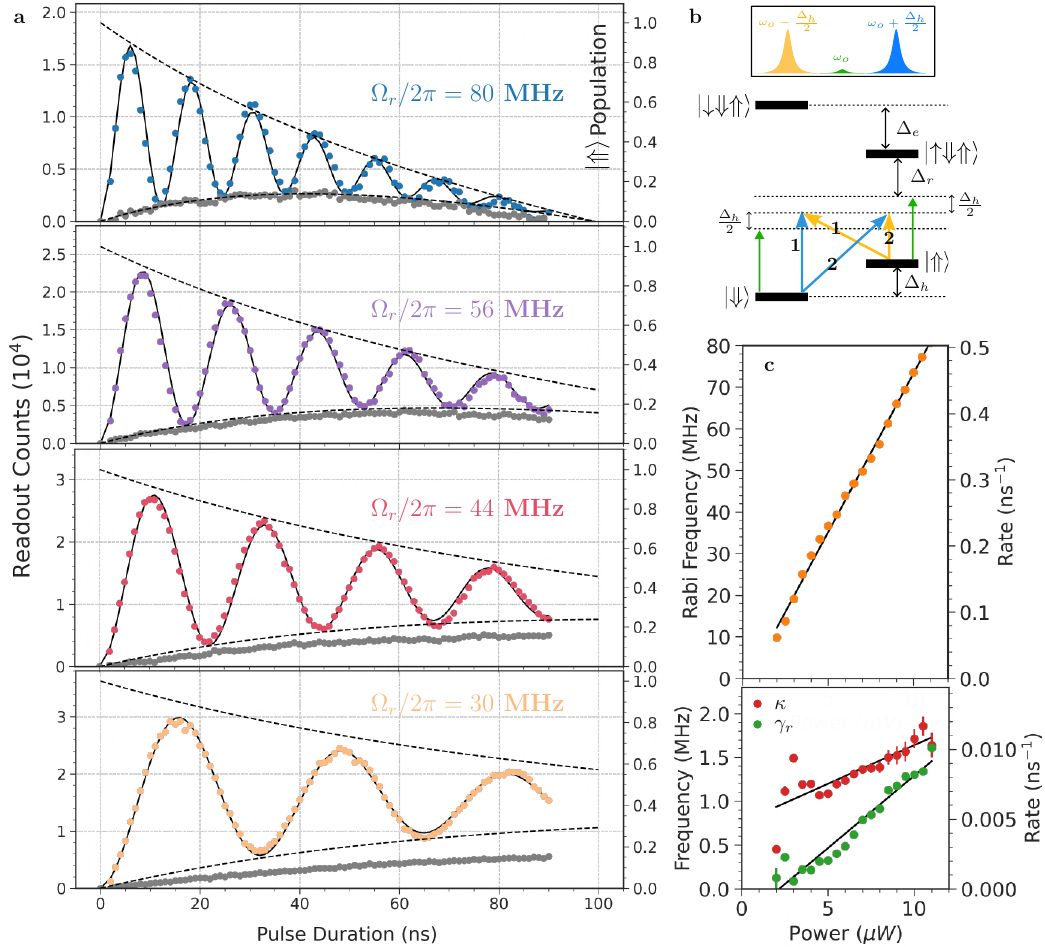}
	\caption{\textbf{Rabi oscillations between the hole spin states.} \textbf{(a)} Measured (dotted) and fitted (black lines) Rabi curves at different Raman laser powers. Dashed curves correspond to the $\kappa$-induced decay envelopes $\frac{1}{2}\eta_r (1\pm e^{-3\kappa T_r/2})$. Grey curves are measured when the sidebands are off-resonant by $\Delta_{\text{MW}}=2\pi\times245$~MHz. \textbf{(b)} Schematics of optical spin control using a bi-chromatic laser assuming $\Delta_{\text{MW}}=0$. \textbf{(c)} Power-dependence of the spin Rabi frequency $\Omega_r$, spin-flip rate $\kappa$ and $\gamma_r$.}
	\label{fig:Rabi_fit}
\end{figure}

Supplementary Figure~\ref{fig:Rabi_fit} shows measured and fitted Rabi oscillations between the hole spin ground states at different optical powers. To avoid populating the trion states, we set the carrier frequency $\omega_o$ of the Raman laser to be $\Delta_r=2\pi\times350$~GHz detuned from the main transition. The experiment begins by photocreating a hole spin with a $830$~nm pulsed laser then optically pumping the hole spin into $\ketDown$, followed by a pulse driven by the Raman laser with a varying duration $T_r$. The state evolution of the hole spin under the effect of Raman pulse is then probed by detecting fluorescence from $\ketUp$ with a readout pulse. Note that a buffer pulse from the Raman laser is added at the end of the sequence, such that the sequence duty cycle and the average optical power of the laser are kept constant when sweeping $T_r$.

The fluorescence signal from the first $50$~ns of the readout pulse can be fitted by the formula~\cite{Appel2021thesis}:
\begin{align}
    I(T_r) = I_0  \times\eta_r\times\bar{\rho}_{\ketUp\bra{\Uparrow}}(T_r),\label{eq:fitRabi}
\end{align}
where $\eta_r\approx(1-\gamma_r T_r)$ is an empirical model describing the reduced readout efficiency with a power-dependent rate $\gamma_r$. Before fitting the data, the photon counts at $T_r=0$~ns are first subtracted to remove background fluorescence from the readout pulse. Each dataset is numerically fitted with Eq.~(\ref{eq:fitRabi}) using $\kappa$, $\gamma_r$, $I_0$ and $\Omega_r$ as free parameters, and with the standard deviation in Overhauser field fluctuations $\sigma_{\text{OH}}$ fixed. The fitted peak intensity $I_0$ can then be extracted to normalize the dataset. For the fit, we take $\sigma_{\text{OH}}=\sqrt{2}/T^*_2\approx2\pi\times9.7$~MHz for a measured spin dephasing time of $T^*_2\approx23.2$~ns~\cite{Appel2021b}. In the limit of $\Omega_r>5\sigma_{\text{OH}}$, the spin noise dynamics is dominated by $\kappa$ as the curves in Supplementary Figure~\ref{fig:Rabi_fit}a become tightly bounded by the decay envelopes $\frac{1}{2}\eta_r (1\pm e^{-3\kappa T_r/2})$.

Supplementary Figure~\ref{fig:Rabi_fit}c shows the fitted spin Rabi frequency $\Omega_r$, the power-dependent spin-flip rate $\kappa$ and $\gamma_r$, which all increase linearly with Raman laser power, as observed in Refs.~\cite{Bodey2019,Appel2021b}. For modelling the entanglement infidelity due to imperfect spin rotations, we need to quantify the fidelity of the spin $\pi$-rotation $F_\pi$. For each normalized Rabi curve in Supplementary Figure~\ref{fig:Rabi_fit}a, $F_\pi$ is estimated from its first peak at which the $\ketUp$ population is maximized. 

In the entanglement protocol experiment, a $T_r=7$~ns $\pi$-pulse (corresponding to $\Omega_r\approx71$~MHz) is used. Therefore, $F_\pi=(88.1\pm3.8)\%$ with $\kappa=(0.0098\pm0.0007)\text{ ns}^{-1}$ and $\gamma_r=(0.0081\pm0.0002)\text{ ns}^{-1}$, with error bounds obtained from the fit. The estimated rotation fidelity is consistent with the one extracted previously on the same QD ($F_\pi=(88.5\pm0.3)\%$)~\cite{Appel2021b}.

\newpage
\section*{Supplementary Note 7: Protocol Performance Metrics}
\label{sec:metrics}
In this section we investigate the metrics that characterize the performance of the entanglement protocol. The protocol should operate at high speed, with a high fidelity, be tolerant to noise, and have low insertion loss. We estimate these metrics for our system to compare with the state-of-the-art photon-scattering protocols on other platforms in Supplementary Table~\ref{table:compare}.
\begin{table}[ht]
    \centering
    \begin{tabularx}{0.9\textwidth}{Y c c *{4}{Y}}
    \toprule
    \multirow{2}{*}{\textbf{Platform}} & \multicolumn{2}{c}{\textbf{Fidelity}} & \multirow{2}{*}{\textbf{Duration}} & \multirow{2}{*}{\textbf{Efficiency}} & \multirow{2}{*}{\textbf{Entanglement rate*}} & \multirow{2}{*}{\textbf{Ref.}}\\ 
    &  Raw & Corrected \\
    \colrule
    SiV-PCC & 0.7 & 0.89-0.94 & 30 $\mu$s & 0.493 & 11.9~Hz &\cite{Nguyen2019b,Bhaskar2020}\\ 
    QD-PCW & 0.66 & 0.74  & 0.29 $\mu$s & 0.171 & 4.75 Hz & Present work\\
    Rb-cavity & $0.86^\dagger$ & -  & 146~$\mu$s & 0.39 & 34.5 Hz & \cite{Kalb2015}\\
    \botrule
    \end{tabularx}
    \caption{\textbf{Comparison of various performance metrics.} PCC: Photonic crystal cavity. PCW: Photonic crystal waveguide. *See Supplementary Table~\ref{table:rate} for estimation details. ${}^\dagger$Not specified if raw or corrected.}
    \label{table:compare}
\end{table} 

From Supplementary Table~\ref{table:compare}, several conclusions can be drawn. First, the raw (corrected) entanglement fidelity $66\%$ ($74\%$) in QD-PCW is comparable to those achieved in SiV-PCC~\cite{Nguyen2019b,Bhaskar2020}. It is noted that the $89\%$ fidelity reported in Ref.~\cite{Nguyen2019b} should be regarded as an inferred maximum fidelity after correcting for imperfect single-shot readout. In contrast our reported $74\%$ fidelity only corrects for subtraction of background fluorescence due to time delay of rotational pulses in the long path of the interferometer (Supplementary Note 9), which can be readily mitigated using shorter pulses and better pulse shaping, or longer interferometric delays.

Second, the protocol duration of the QD-PCW is two-orders of magnitude shorter than that of SiV-PCC. This is primarily due to the long duration ($30~\mu s$) required to distinguish between the two SiV spin states during single-shot readout~\cite{Bhaskar2020}. Similarly the protocol duration is even longer in Rb-cavity system. Supplementary Table~\ref{table:duration} shows a breakdown of the protocol duration, which is  defined as the total duration of all pulses required to run the entanglement sequence. This includes the spin initialization pulse, two optical pulses with a delay that made up the photonic qubit, and the spin rotation $\pi/2$- and $\pi$-pulses. For both systems, the emitter's lifetime is short compared to the duration of each photonic pulse, and therefore not included in the protocol duration. A faster operating speed without compensating for entanglement fidelity therefore enables quantum communication at a higher clock rate.
\begin{table}[ht]
    \centering
    \begin{tabularx}{0.85\textwidth}{*{7}{Y}}
    \toprule
    \textbf{Platform} & \textbf{Initialization} & \textbf{$2T_p$} &  \textbf{TBI delay} & \textbf{$\pi/2$-pulse} & \textbf{$\pi$-pulse} & \textbf{Ref.} \\ 
    \colrule
    SiV-PCC & $30~\mu$s & 20~ns & 142~ns & 16~ns & 32~ns & \cite{Bhaskar2020}\\
    SiV-PCC & $13~\mu$s & 10~ns & 30~ns & 6~ns & 12~ns & \cite{Nguyen2019b}\\
    QD-PCW & $0.2~\mu$s & 4~ns & 11.8~ns & 3.5~ns & 7~ns & Present work\\
    Rb-cavity & $140$~$\mu$s & 0.6~$\mu$s* & - & 1.7~$\mu$s & 3.4~$\mu$s & \cite{Kalb2015}\\
    \botrule
    \end{tabularx}
    \caption{\textbf{Breakdown of the total protocol duration in each step.} $T_p$: duration of a photonic pulse. TBI: time-bin interferometer. *The Rb-cavity scheme uses polarization-encoding for the photonic qubit.}
    \label{table:duration}
\end{table}

For both schemes, the entanglement generation is conditioned on detecting a reflected photon, and is therefore ideally successful $50\%$ of the time in each run since at most half of the input photon gets reflected when the spin is initialized in a superposition state. This means that the protocol success probability $P_s$ depends on the spin-dependent reflectivity of the emitter, which can be expressed by Eq.~(\ref{eq:tr}):
\begin{align}
    P_s &= \frac{1}{2}\bigg[\int^{\infty}_{-\infty} \abs{\Phi_1(\omega)}^2\abs{r_1(\omega)}^2d\omega + \int^{\infty}_{-\infty} \abs{\Phi_1(\omega)}^2\abs{\mathring{r}_1(\omega)}^2d\omega+P^{\omega_1}_{\gamma_d}\bigg].
\end{align}
We find that for each incoming photon, it is reflected with a probability of $33.9\%$ from the $\ketUp$ state and with a probability of $0.05\%$ from $\ketDown$. The protocol efficiency is therefore $(17.1\pm0.3)\%$. The  reflectivity is inferred from the two-color transmission experiment~(Supplementary Note 4), which depends on the device cooperativity and noise properties. 

Another conclusion drawn from Supplementary Table~\ref{table:compare} is that the entanglement rates of the QD-PCW and SiV-PCC systems are of similar order of magnitude despite the difference in efficiency. In our work, heralded spin readout is used instead since the single-shot fidelity could only reach $52\%$ due to limited collection efficiency $\eta_c\approx0.3\%$ and cyclicity $C=14.7$~\cite{Appel2021b}. The collection efficiency can readily be improved in next-generation devices. The low coupling rate is compensated by the fast repetition rate of the system. As opposed to the $200~$ms long calibration step, which includes active locking of the TBI and preselection procedures used to lock the SiV resonance, our experiment is able to be executed in $606$~ns. The self-stabilizing TBI in this work does not require active locking and offers week-long stability with $>99\%$ interferometric visibility. 
\begin{table}[ht]
    \centering
    \begin{tabularx}{0.9\textwidth}{*{5}{Y}}
    \toprule
    \textbf{Platform} & \textbf{$P_{\text{photon}}$ (\%)} & \textbf{$P_{\text{spin}}$ (\%)} &  \textbf{$\tau_{\text{seq}}$} & \textbf{Entanglement rate}\\ 
    \colrule
    SiV-PCC & $2\langle n \rangle_m \eta=0.17$ & $F_{\text{ssr}}\eta=42.3$ & 60 $\mu$s & 11.9 Hz \\
    QD-PCW & $0.027$ & $1.1$ & $0.606~\mu$s & 4.75 Hz \\
    Rb-cavity & $\langle n \rangle_m \eta=3.45$ & $100$ & $0.001~$s & 34.5 Hz \\
    \botrule
    \end{tabularx}
    \caption{\textbf{Entanglement rate estimation.} $P_{\text{photon}}$ ($P_{\text{spin}}$): probability of detecting a reflected photon during readout of the photonic (spin) qubit in a single run. $\tau_{\text{seq}}$: sequence duration including spin initialization, protocol sequence and state readout. $\langle n \rangle_m$: average number of photon per pulse incident on the cavity. $\eta$: total heralding efficiency. $F_{\text{ssr}}$: fidelity of single-shot readout. For QD-PCW, $P_{\text{photon}}$ ($P_{\text{spin}}$) is obtained by summing all photon clicks during the photonic (spin) detection windows, then divided by the 100~s integration time and repetition rate. For Rb-cavity, $\eta\equiv \mathcal{R}\eta_{\text{det}}=0.69\times0.56=0.38$ and $\langle n \rangle_m=0.09$.}
    \label{table:rate}
\end{table}

In Supplementary Table~\ref{table:rate} we estimate the entanglement rates of both systems using $\mathcal{R}\equiv P_{\text{photon}}\times P_{\text{spin}}\times \mathcal{R}_{\text{rep}}$ where the experimental repetition rate is given by $\mathcal{R}_{\text{rep}}=1/\tau_{\text{seq}}$. $P_{\text{photon}}$ ($P_{\text{spin}}$) is the probability of detecting a reflected photon in one iteration of the experiment during the photon (spin) readout window. For a fair comparison, we define $\tau_{\text{seq}}$ to be the time it takes to initialize, generate entanglement and read out the final state, rather than the total sequence time that includes calibration steps. To validate our rate estimation of $4.75~$Hz, we predict an entanglement rate of $4.75~\text{Hz}\times 4/(2\bar{n})\approx 106\pm15~$Hz for Bell-state generation using optical excitations, where 2$\bar{n}=2\times(0.089\pm0.012)$ is the total number of photons scattered off the QD and the factor of $4$ originates from the $50\%$ reduction in waveguide collection efficiency due to the cross-polarized scheme (Supplementary Note 1). This is in excellent agreement with $124$~Hz extracted rate in Ref.~\cite{Appel2021b}. A $25$-fold improvement in the entanglement rate is possible as near-$90\%$ single-shot readout fidelity on a sub-nanosecond timescale with collection efficiency $\eta_c=76\%$ can be  achieved~\cite{Antoniadis2022}. To reach an even higher entanglement rate, it is necessary to improve the device reflectivity by reducing the spectral diffusion noise and increasing the QD-waveguide coupling efficiency $\beta$. 

\section*{Supplementary Note 8: Concurrence Estimate}
\label{sec:concurrence}
Another measure frequently used in the literature to quantify entanglement is the concurrence $\mathcal{C}$, which for a non-separate bipartitle system $\mathcal{C}>0$ and reaches unity for a maximally entangled state. Here we estimate a lower bound for the concurrence using the raw and background-corrected coincidences recorded in the experiment (shown in Figures~3b-c of the main text). The concurrence is given by
\begin{align}
    \mathcal{C}=\text{max}(0,\sqrt{\lambda_0}-\sum^{N}_{i-1}\sqrt{\lambda_i}),\label{eq:concurrence}
\end{align}
where $\lambda_i$ are eigenvalues of the matrix $\rho_{\text{meas}}(\hat{\sigma}^{(p)}_y\otimes \hat{\sigma}^{(s)}_y)\rho_{\text{meas}}(\hat{\sigma}^{(p)}_y\otimes \hat{\sigma}^{(s)}_y)^\dagger$, and $\rho_{\text{meas}}$ is the measured normalized density matrix of the spin-photon state assuming negligible off-diagonal entries~\cite{Nguyen2019b}
\begin{align}
    \rho_{\text{meas}}=\left[\begin{array}{cccc}
    \rho_{e\Uparrow,e\Uparrow} & 0 & 0 & 0\\
    0 & \rho_{e\Downarrow,e\Downarrow} & \rho_{l\Uparrow,e\Downarrow} & 0\\
    0 & \rho_{e\Downarrow,l\Uparrow} & \rho_{l\Uparrow,l\Uparrow} & 0\\
    0 & 0 & 0 & \rho_{l\Downarrow,l\Downarrow}
    \end{array}\right],
\end{align}
which shares the same form as Eq.~(\ref{eq:spinflipfinal}). Using the coincidence counts in Figures~3b-c of the main text and Eq.~(\ref{eq:concurrence}), we find a raw (background-corrected) concurrence of $\mathcal{C}\geq0.345 \pm0.012$ ($0.495\pm0.105$) with error bounds obtained by Monte Carlo simulations.

\section*{Supplementary Note 9: Fidelity Background Subtraction}
\label{sec:bg}
Background photons in this work include leakage from the photonic qubit and readout lasers, detector dark counts, accidental coincidence counts, and fluorescence from the rotational laser. Prior to running the experiment, both the optical paths for the photonic qubit and spin readout are optimized such that the signal-to-noise ratio is $>100$. The photon detection uses superconducting nanowire single-photon detectors
(SNSPDs) with a dark count rate of $10~$Hz. False coincidence counts account for $<3\%$ of all detected coincidences. A majority of the background subtraction is done by removing the fluorescence during the rotational pulses which overlaps in time with the photon detection window, see inset of Supplementary Figure~\ref{fig:RotationOverlay}. The fluorescence may be due to photo-ionization by the red-detuned Raman laser~\cite{Lochner2021} followed by subsequent reinjection of a hole, which leads to photon emission through the unfiltered cycling transition. This hypothesis could be supported by the fact that a $40~$ns buffer pulse at 540~ns is used partly to inject a hole in the current device. The fluorescence from these pulses thus enters the detection path of the TBI and half of which is then delayed in time, partially overlapping with the time-bin detection windows.
\begin{figure}
     \centering
     \includegraphics[scale=0.8]{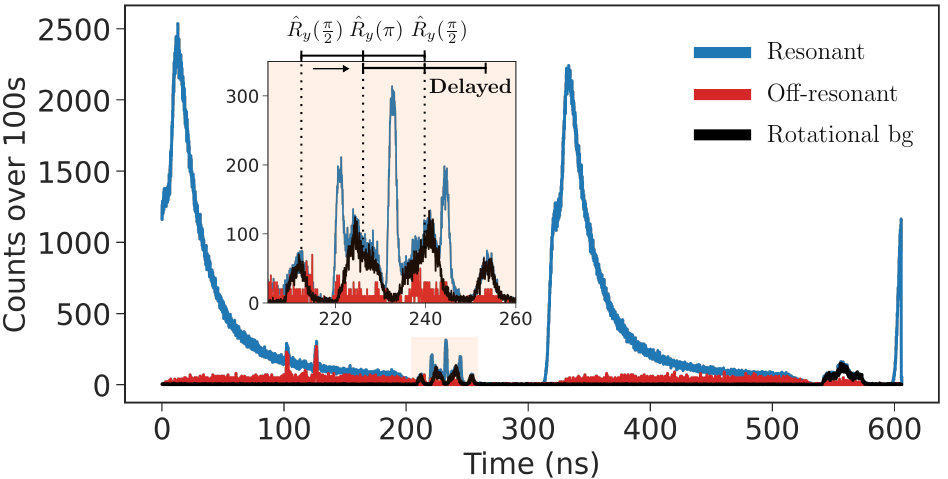}
     \caption{\textbf{Entanglement pulse sequence overlaid with rotational background fluorescence.} Fluorescence from the 350~GHz off-detuned rotational pulses near 200~ns can be seen in the time-resolved histogram recorded by the detectors. The inset shows a magnified view of the pulse sequence, where the background fluorescence from the rotational laser partially overlaps in time with the photon detection.  }\label{fig:RotationOverlay}
\end{figure}


\bibliographystyle{naturemag}
\bibliography{reflist}
